\documentclass[usenatbib]{mn2e}

\input epsf

\usepackage{graphicx}
\usepackage{txfonts}
\usepackage{epsfig}

\title[Super-Nyquist asteroseismology]{Super-Nyquist asteroseismology
  of solar-like oscillators with \emph{Kepler} and K2---expanding the
  asteroseismic cohort at the base of the red-giant branch}

\author[Chaplin et al.]{W.~J.~Chaplin$^{1,2}$\thanks{E-mail:
    w.j.chaplin@bham.ac.uk}, Y.~Elsworth$^{1,2}$,
  G.~R.~Davies$^{1,2}$, T.~L.~Campante$^{1,2}$,
  R.~Handberg$^{1,2}$,\newauthor A.~Miglio$^{1,2}$ and
  S.~Basu$^{3}$\\ $^1$ School of Physics \& Astronomy, University of
  Birmingham, Edgbaston, Birmingham, B15 2TT, UK\\ $^2$ Stellar
  Astrophysics Centre (SAC), Department of Physics and Astronomy,
  Aarhus University, Ny Munkegade 120, DK-8000 Aarhus C,
  Denmark\\ $^3$ Department of Physics and Astronomy, Yale University,
  P.O. Box 208101, New Haven, CT, 06520, USA}

\begin{document}

\maketitle

\begin{abstract}

We consider the prospects for detecting solar-like oscillations in the
``super-Nyquist'' regime of long-cadence (LC) \emph{Kepler}
photometry, i.e., \emph{above} the associated Nyquist frequency of
$\simeq 283\,\rm \mu Hz$. Targets of interest are cool, evolved
subgiants and stars lying at the base of the red-giant branch. These
stars would ordinarily be studied using the short-cadence (SC) data,
since the associated SC Nyquist frequency lies well above the
frequencies of the detectable oscillations. However, the number of
available SC target slots is quite limited. This imposes a severe
restriction on the size of the ensemble available for SC asteroseismic
study.  We find that archival \emph{Kepler} LC data from the nominal
Mission may be utilized for asteroseismic studies of targets whose
dominant oscillation frequencies lie as high as $\simeq 500\,\rm \mu
Hz$, i.e., about 1.75-times the LC Nyquist frequency. The frequency
detection threshold for the shorter-duration science campaigns of the
re-purposed \emph{Kepler} Mission, K2, is lower. The maximum threshold
will probably lie somewhere between $\simeq 400$ and $450\,\rm \mu
Hz$. The potential to exploit the archival \emph{Kepler} and K2 LC
data in this manner opens the door to increasing significantly the
number of subgiant and low-luminosity red-giant targets amenable to
asteroseismic analysis, overcoming target limitations imposed by the
small number of SC slots.  We estimate that around 400 such targets
are now available for study in the \emph{Kepler} LC archive. That
number could potentially be a lot higher for K2, since there will be a
new target list for each of its campaigns.

\end{abstract}


 \section{Introduction}
 \label{sec:intro}

The NASA \emph{Kepler} Mission has provided photometric observations
of exquisite quality for asteroseismic studies of a diverse range of
pulsating stars (Gilliland et al. 2010). Particularly noteworthy has
been the large volume of data collected on cool main-sequence,
subgiant and red-giant stars showing solar-like oscillations,
pulsations that are excited and intrinsically damped by near-surface
convection (Chaplin \& Miglio 2013). The successful asteroseismology
programme looks set to continue in the re-purposed \emph{Kepler}
Mission, K2 (Howell et al. 2014). The quality of the K2 photometry for
asteroseismic studies of coherent pulsators has already been
demonstrated by results from engineering test data (Jeffery \& Ramsay
2014; Hermes et al. 2014). The photometric performance has also
confirmed the predicted potential for studies of solar-like
oscillators (Chaplin et al. 2013), studies that will be made possible
by the longer, full-campaign science data.

\emph{Kepler} data were collected in two cadences (Koch et al. 2010)
during the 4-yr-long nominal Mission. Data on up to 170,000 targets
were available in the 29.4-minute long cadence (LC), whilst a much
smaller total of up to 512 targets could be observed simultaneously in
the 58.85-second short cadence (SC).  These cadences establish Nyquist
frequencies for the two modes of operation of $\simeq 283\,\rm \mu Hz$
(in LC) and $\simeq 8496\,\rm \mu Hz$ (in SC). The LC Nyquist
frequency coincides with the dominant periods of oscillation shown by
cool stars that lie just above the base of the red-giant branch. Both
cadences have been maintained for the K2 Mission, but the total number
of targets for each observing campaign has been reduced -- to
approximately 10,000 and 50 targets, respectively -- because of the
need for larger pixel target masks.

The detection of solar-like oscillations in cool main-sequence and
subgiant stars, and stars at the very bottom of the red-giant branch,
demands that the target-limited SC data be utilized since the dominant
periods of oscillation are shorter than one hour. Observations made in
the numerous long-cadence slots are sufficient to detect oscillations
in more evolved red giants, because the relevant pulsation periods are
longer.

Oscillations have already been detected in around 16,000 red giants in
the nominal Mission LC data (Hekker et al. 2011; Stello et al. 2013;
Huber et al. 2014). Whilst \emph{Kepler} increased the number of cool
main-sequence and subgiant stars with asteroseismic data by over an
order of magnitude, to around 700 targets (Chaplin et al., 2011a,
2014; Huber et al. 2013), totals were nevertheless limited by the
available number of SC target slots. The possibility of detecting
solar-like oscillations in LC data \emph{above} the LC Nyquist
frequency -- i.e., in the ``super-Nyquist'' regime -- offers the
potential to increase significantly the number of evolved subgiant and
low-luminosity red-giant targets amenable to asteroseismic study.

The use of LC data for super-Nyquist studies of coherent pulsators has
been considered in some detail by Murphy et al. (2013). Baran et
al. (2012) also employed a super-Nyquist analysis of a compact
pulsator observed in SC. While Gaulme et al. (2013) and Beck et
al. (2014) identified a few red giants with oscillation frequencies
above the LC Nyquist frequency, the prospects for studying solar-like
oscillators in the LC super-Nyquist regime have not yet been explored
in any detail. Our aim in this paper is to consider the utility of the
archival \emph{Kepler} and future K2 data for such studies.

The layout of the rest of the paper is as follows. We begin in
Section~\ref{sec:recap} with a summary of some of the basic principles
from Fourier analysis that are relevant to our study. We then consider
in Section~\ref{sec:kepler} the specific case of \emph{Kepler}
long-cadence observations; and in Section~\ref{sec:solar} we go on to
consider the characteristics of the super-Nyquist spectrum shown by
solar-like oscillators, considering the impact of the finite mode
lifetimes (Section~\ref{sec:life}), the use of a priori information to
discriminate true from aliased peaks in the frequency spectrum
(Section~\ref{sec:prior}), and the reduced S/N above the Nyquist
frequency (Section~\ref{sec:sinc}). We include model predictions for
subgiant and low-luminosity red giant stars, and real super-Nyquist
examples from the \emph{Kepler} archive (Section~\ref{sec:real}). We
finish the paper in Section~\ref{sec:conc} with concluding remarks.

 \section{Basic principles}
 \label{sec:recap}

Let us begin by considering the simple case of a sinusoidal signal of
frequency $\nu$, sampled at regular intervals $\Delta t$ in time.
This establishes a sampling frequency $\nu_{\rm s} = (\Delta t)^{-1}$.
The well-known sampling theorem (Nyquist 1928; Shannon 1949) tells us
that when $\nu \le \nu_{\rm s}/2$, the sampling is sufficient to
completely determine the signal -- we say the signal is oversampled --
and there is no ambiguity in the measured frequency. We may also write
the sampling requirement as $\nu \le \nu_{\rm Nyq}$, where the Nyquist
frequency, $\nu_{\rm Nyq}$, is defined as
 \begin{equation}
 \nu_{\rm Nyq} \stackrel{\mathrm{def}}{=} \nu_{\rm s}/2 \equiv (2\Delta t)^{-1}.
 \label{eq:nyq}
 \end{equation}
What happens if instead the signal is undersampled, so that $\nu >
\nu_{\rm Nyq}$? Power will be present in the ``super-Nyquist'' regime
of the frequency spectrum at the true frequency, $\nu$; and it will
also now be reflected, or aliased, back into the frequency region
below $\nu_{\rm Nyq}$. If we write the true frequency as $\nu =
\nu_{\rm Nyq} + \nu'$, there will be peaks at $\nu_{\rm Nyq} + \nu'$
and $\nu_{\rm Nyq} - \nu'$.  There is now an ambiguity in the
estimated frequency, and we are completely reliant on using other
knowledge to pick the true frequency. We shall see that irregular time
sampling offers the potential to lift this degeneracy provided the
signal is coherent, or nearly coherent, over the duration of the
observations (of which more in Section~\ref{sec:kepler} below).

The appearence of the frequency spectrum is also affected by the
amount of time during each cadence $\Delta t$ that is used to collect
data, i.e., in our case, to integrate photons from the target star.
If a high fraction of each cadence is used to collect data, we may
significantly underestimate the true amplitude because each datum may
average the time-varying signal.  If the integration time per cadence
is $\Delta t'$, then a signal of frequency $\nu$ will have its
amplitude attenuated by the factor (e.g., Campante 2012):
 \begin{equation}
 \eta = {\rm sinc} \left[ \pi \left( \nu \Delta t' \right) \right].
 \label{eq:sincgeneral}
 \end{equation}
When the fraction of each cadence given over to integration is unity,
so that $\Delta t' = \Delta t$, the attenuation may then be written as
(e.g., Chaplin et al. 2011b; Huber et al. 2011; Murphy 2012):
 \[
 \eta = {\rm sinc} \left[ \pi \left( \nu \Delta t \right) \right] =
      {\rm sinc} \left[ \pi \left( \frac{\nu}{\nu_{\rm s}} \right)
        \right]
 \]
 \begin{equation}
 ~~~~~~~~~~~~~~~~~~~~~~~~~~~~\equiv {\rm sinc} \left[ \pi/2 \left(
     \frac{\nu}{\nu_{\rm Nyq}} \right) \right].
 \label{eq:sinc}
 \end{equation}
The attenuation in power is given by the square of the sinc function,
$\eta^2$. Even when the integration duty cycle is close to 100\,per
cent and the attenuation is at its strongest -- which is the case for
\emph{Kepler}; see Section~\ref{sec:kepler} below --
Equation~\ref{eq:sinc} indicates that one still has sensitivity to
signals in the super-Nyquist regime ($\nu > \nu_{\rm Nyq}$) since the
first zero of the sinc function does not occur until $\nu = \nu_{\rm
  s} \equiv 2\nu_{\rm Nyq}$. It is also worth remarking that even when
signals are oversampled, so that $\nu \le \nu_{\rm Nyq}$, there is
still significant attenuation close to the Nyquist frequency.

 \section{\emph{Kepler} long-cadence observations}
 \label{sec:kepler}


\begin{figure*}
 \centerline {\epsfxsize=9.0cm\epsfbox{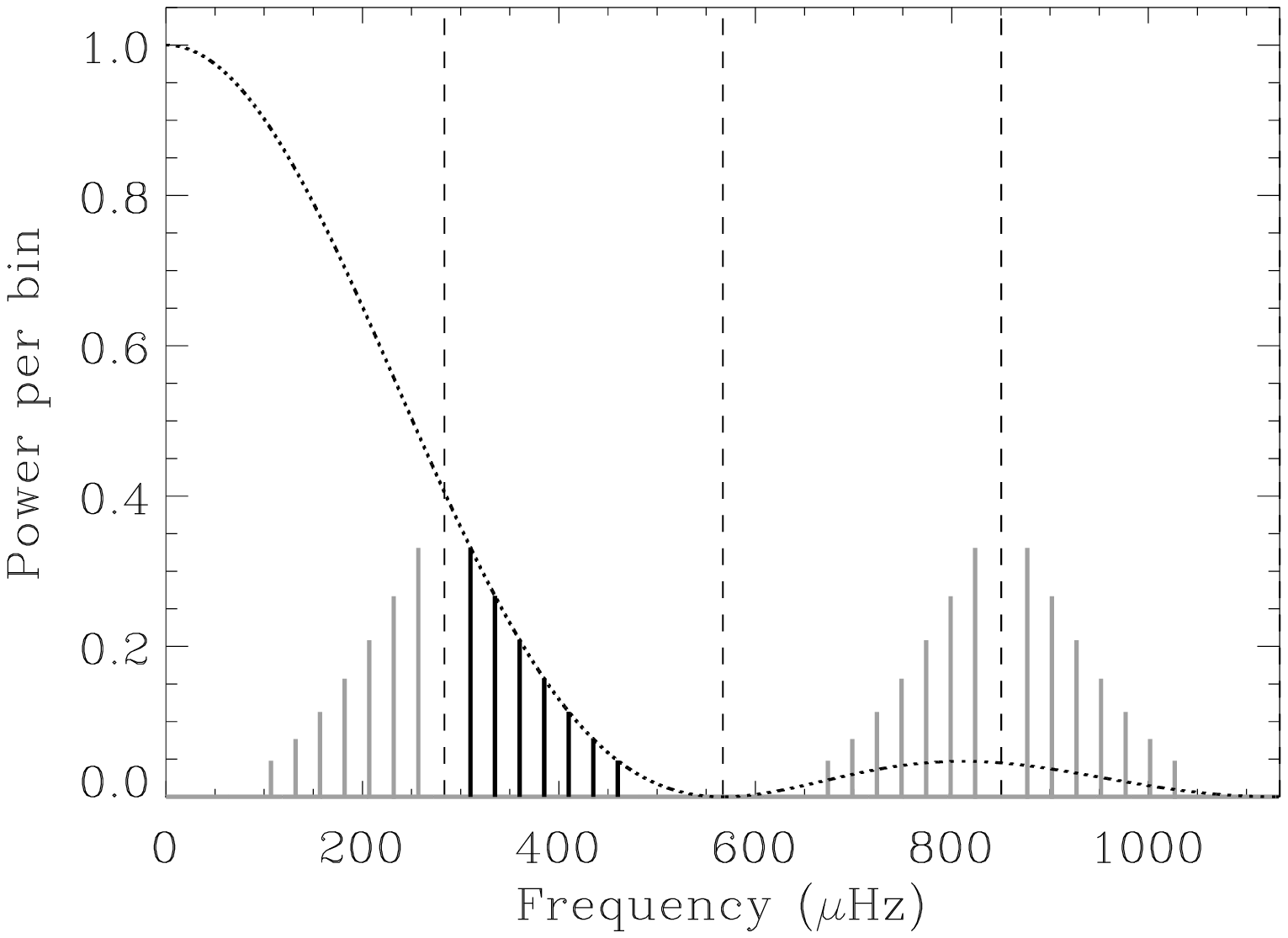}
              \epsfxsize=9.0cm\epsfbox{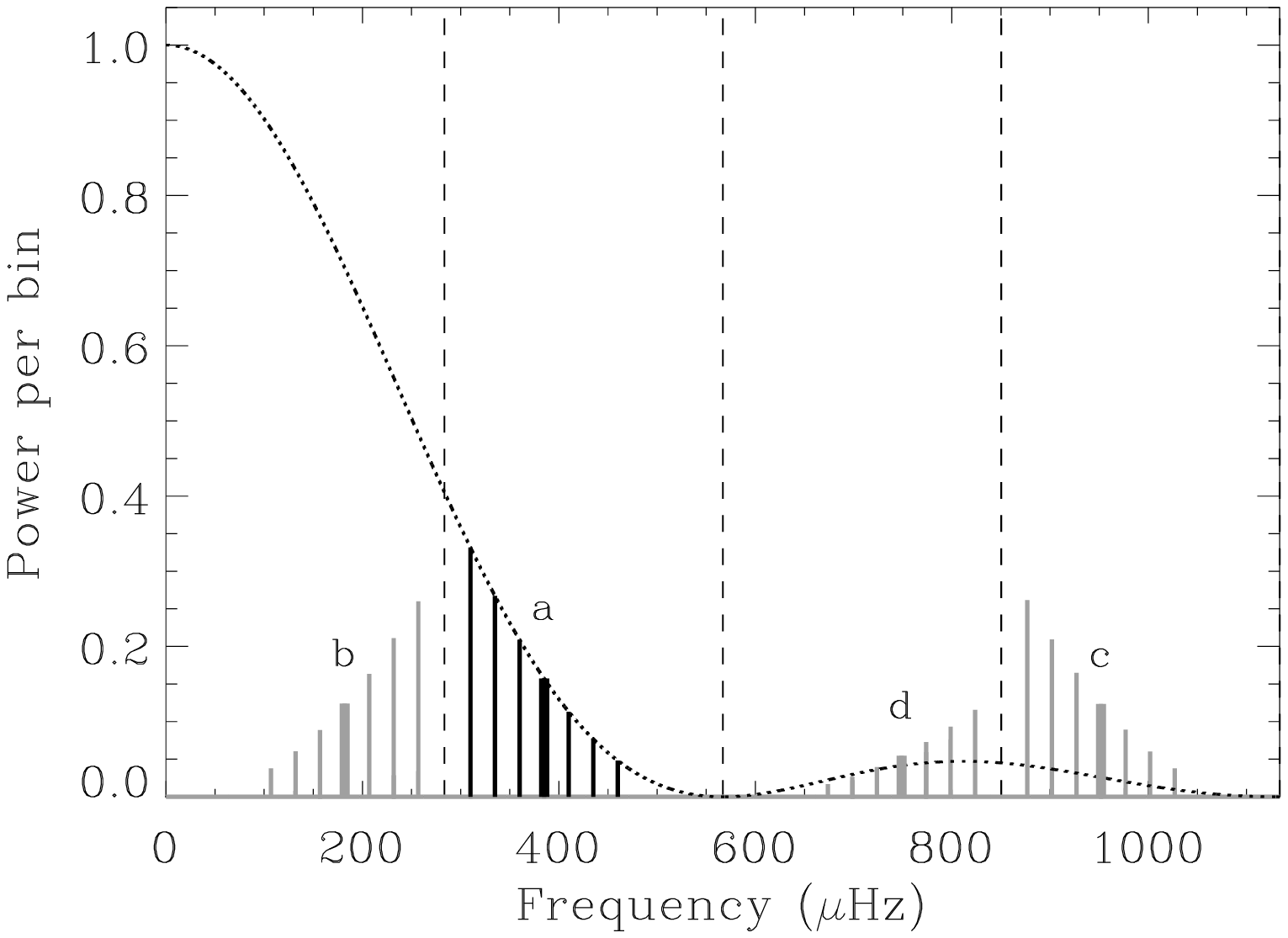}}

 \caption{\small Left-hand panel: Frequency-power spectrum for a
   series of sinusoids of unit amplitude having frequencies between
   $310\,\rm \mu Hz$ and $460\,\rm \mu Hz$, sampled on a regular
   cadence. Peaks due to the true frequencies are rendered as black
   lines. The vertical dashed lines mark multiples of $\nu_{\rm
     Nyq}$. Peaks in grey are aliases of the true frequencies. The
   dotted line marks the sinc-squared attenuation envelope (i.e.,
   $\eta^2$).  Right-hand panel: Spectrum now given by irregular
   sampling of the undersampled sinusoids. The true peak and the
   aliases of one of the frequencies are rendered in thick linestyles
   (see text).}

 \label{fig:fig1}
\end{figure*}


The \emph{Kepler} LC data are comprised of $\Delta t=29.4$-min
cadences (Jenkins et al. 2010) that are exactly regular in the
spacecraft frame of reference. This establishes a notional LC Nyquist
frequency of $\nu_{\rm Nyq} \simeq 283\,\rm \mu Hz$.  Each 29.4-min
cadence is in turn a summation of 270 individual $\simeq 6$-sec
readouts (e.g., Gilliland et al. 2011). Most of each cadence is
therefore given over to the collection of photons. The very high
fractional duty cycle for the \emph{Kepler} integrations means the
signal attenuation follows closely that described by
Equation~\ref{eq:sinc}.

The left-hand panel of Fig.~\ref{fig:fig1} shows the frequency-power
spectrum of idealized observations, made on a regular
\emph{Kepler}-like cadence, of several undersampled sinusoids. The
sinudoids are all of unit amplitude and have frequencies lying in the
super-Nyquist regime between $310\,\rm \mu Hz$ and $460\,\rm \mu Hz$.
Peaks due to the true frequencies are rendered as black lines. The
vertical dashed lines mark multiples of $\nu_{\rm Nyq}$. Peaks in grey
are therefore aliases of the true frequencies.  The spectrum is
repeated every $2\nu_{\rm Nyq}$ because of the discrete nature of the
calculation.

This power spectrum has been calibrated\footnote{Here, and throughout
  the rest of the paper, we show spectra calibrated in power per bin
  or power per Hz -- as opposed to amplitude per root bin or per root
  Hz -- since this is the usual approach when analysing solar-like
  oscillators (which we will come to later in Section~\ref{sec:solar}
  below).} so that a sinusoid of unit amplitude with an infinite
sample rate would show a maximum power per bin of unity.  The observed
powers at the true frequencies are much lower than unity, the true
power. This is due to the sinc-function attenuation described by
Equation~\ref{eq:sinc}. The dotted line marks the sinc-squared
suppression envelope in power, i.e., $\eta^2$.

The above is not quite the whole story as far as timing issues for the
\emph{Kepler} observations are concerned. \emph{Kepler} lies in a
372.5-day heliocentric, Earth-trailing orbit. Observations of the
pulsations of a \emph{Kepler} target -- be it one in the original
field, or in the K2 fields in the ecliptic plane -- will be phase
modulated in the spacecraft frame of reference. This is because there
is a component of the orbital motion of \emph{Kepler} along the
line-of-sight (target) direction, which delays or advances the arrival
time of light from the star. The annual size of the effect is
approximately $\pm 190$\,sec for targets in the original field, and
approximately $\pm 500$\,sec for targets in the ecliptic (where the K2
fields lie).

To compensate for this effect\footnote{Aside from the varying
  component due to the orbital motion of the spacecraft about the Sun,
  we assume that there are no other line-of-sight components showing
  significant variation on the timescale of the observations, e.g.,
  variations due to the target being in a short-period binary (see
  Davies et al. (2014) for discussion on issues relating to the
  line-of-sight component).} the \emph{Kepler} time stamps are
corrected to Barycentric arrival times (actually Barycentric Dynamical
Time; see, e.g., Garc\'ia et al. 2014). An important consequence is
that the intervals between time stamps are no longer regular, but are
modulated periodically on a $\sim 1$-yr timescale. This periodic
modulation splits the aliased peaks in the frequency spectrum into
many components, as discussed in detail by Murphy et al. (2013).

The right-hand panel of Fig.~\ref{fig:fig1} shows the result of
simulating LC observations of an idealized star in the original
\emph{Kepler} field undergoing coherent pulsations. The artificial
data span four simulated years and include the aforementioned timing
effects.  The simulated pulsations are the same, undersampled
sinusoids from the left-hand panel. Notice how the aliases are no
longer exact, reflected copies about multiples of $\nu_{\rm
  Nyq}$. Some copies have different maximum power spectral densities
(i.e., peak heights in the spectrum) than others. The reason for this
is that the splitting of the power into several components reduces the
maximum heights in Fig.~\ref{fig:fig1}, relative to those expected for
simple reflected aliases (the total integrated power in the aliases is
conserved).

\begin{figure*}
 \centerline {\epsfxsize=9.0cm\epsfbox{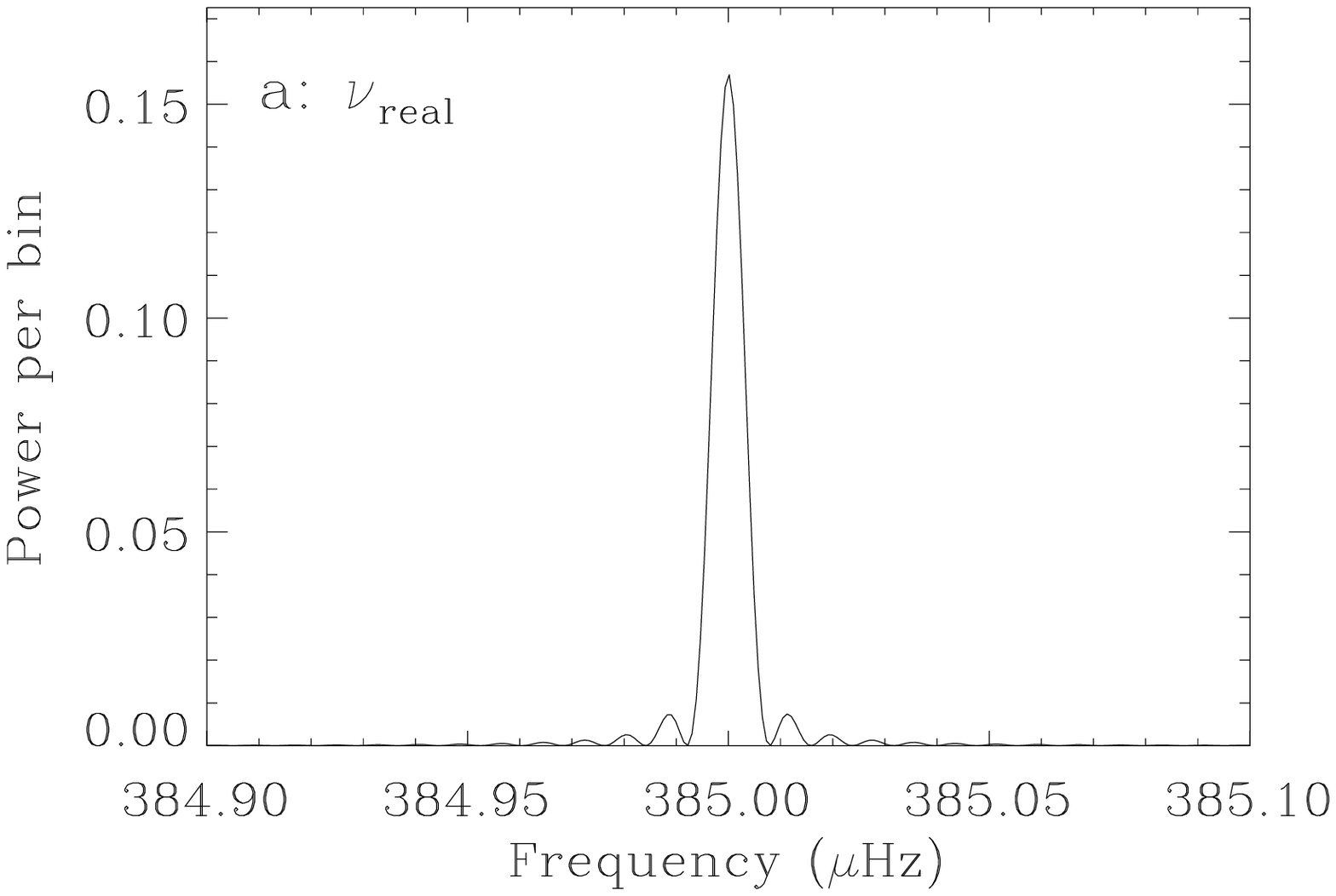}
              \epsfxsize=9.0cm\epsfbox{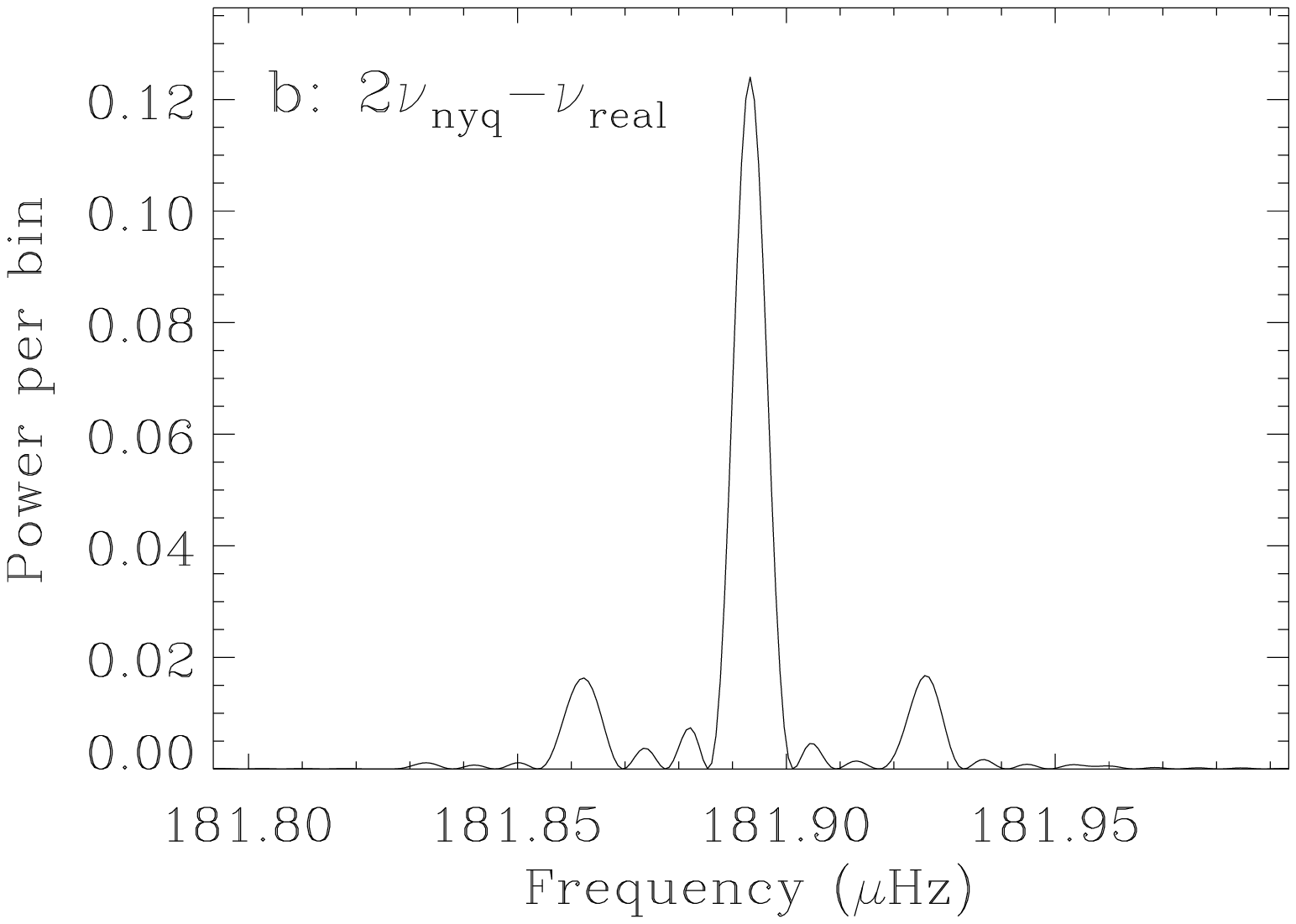}}
 \centerline {\epsfxsize=9.0cm\epsfbox{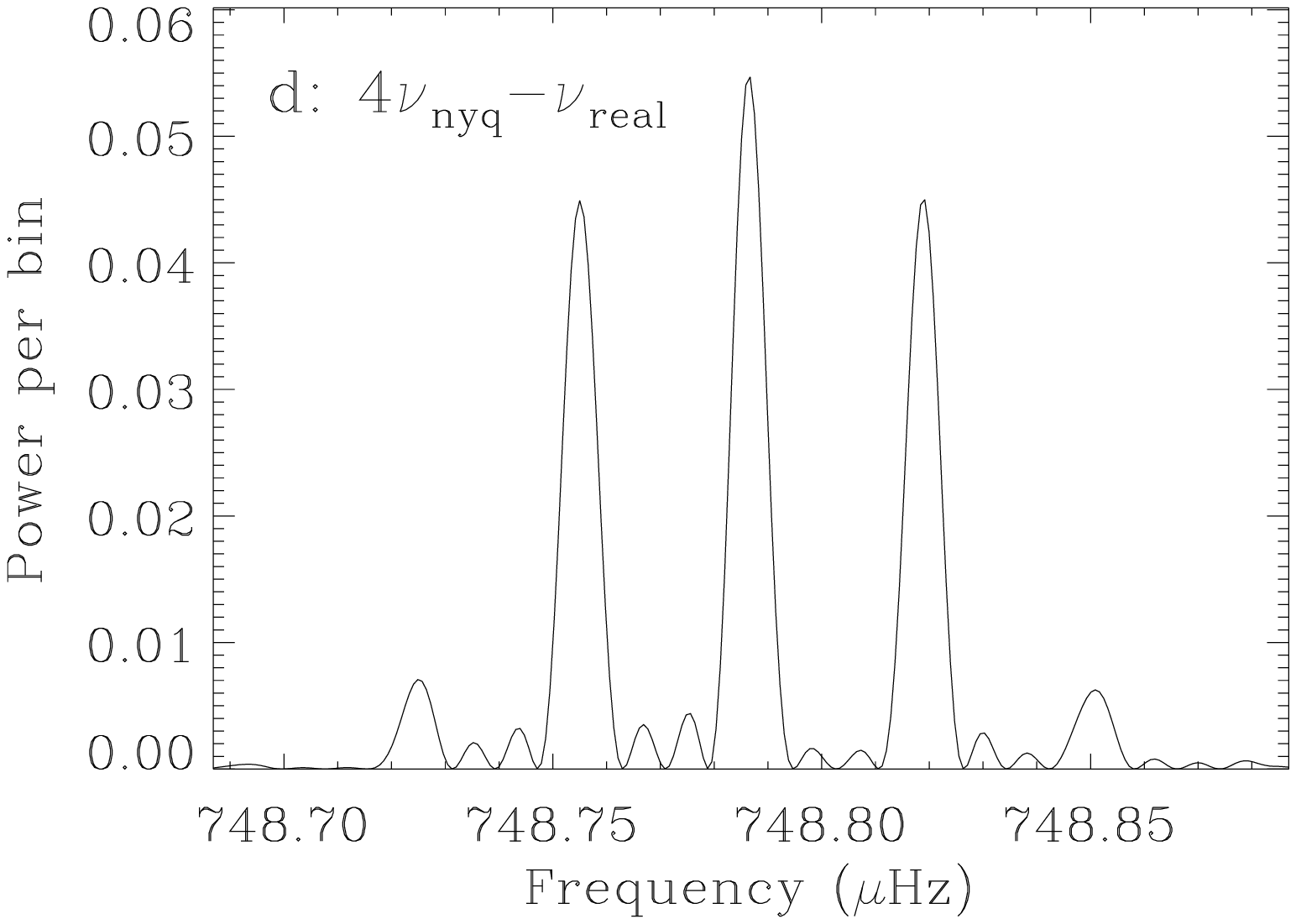}
              \epsfxsize=9.0cm\epsfbox{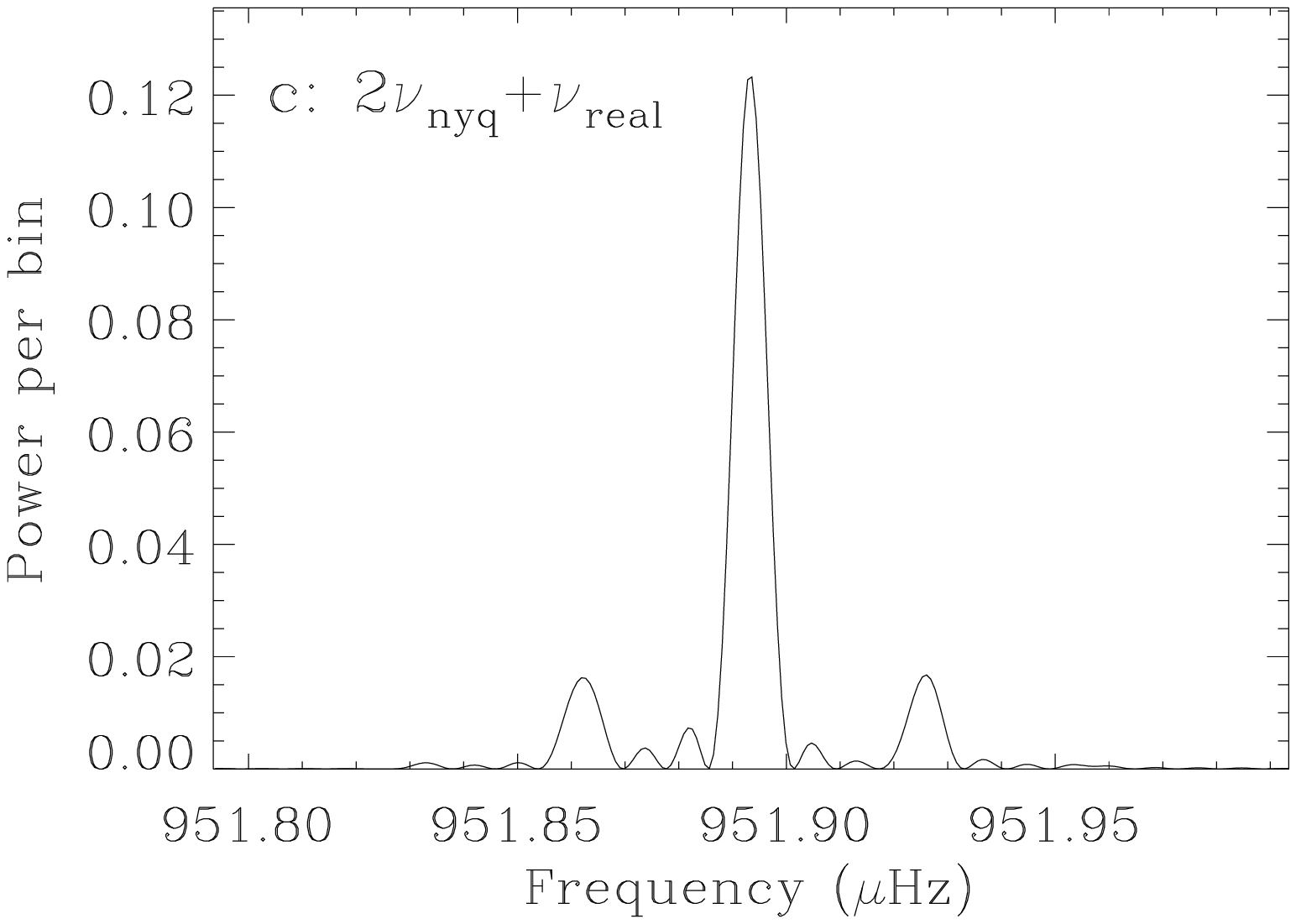}}

 \caption{\small Top left-hand panel: zoom of the peak due to the true
   frequency marked `a' on the right-hand panel
   Fig.~\ref{fig:fig1}. Other panels: zooms of the aliases marked `b',
   `c' and `d' on Fig.~\ref{fig:fig1}.}

 \label{fig:fig2}
\end{figure*}

The top left-hand panel of Fig.~\ref{fig:fig2} shows a zoom of the
peak due to the true frequency marked `a' in the right-hand panel of
Fig.~\ref{fig:fig1}, whilst the other panels show zooms its aliases
marked `b', `c' and `d' on Fig.~\ref{fig:fig1}.  As explained by
Murphy et al. (2013), the exact manner in which the alias peaks are
affected by the timing modulation depends on the relation of their
frequencies to the sampling frequency. If the true frequency is $\nu$,
then aliases at $n \nu_{\rm s} \pm \nu$, or equivalently $2n\nu_{\rm
  Nyq} \pm \nu$, will share the same structure, e.g., what is
predominantly a triplet structure when $n=1$, or a quintuplet
structure when $n=2$, with the frequency splitting between adjacent
components being $\sim 1\,\rm yr^{-1}$.

Murphy et al. (2013) pointed out that for high-amplitude, coherent
pulsations the introduction of sideband structure, which is well
resolved in the nominal Mission \emph{Kepler} data, allows one to
discriminate the real and aliased peaks.  What about solar-like
oscillations, which are not coherent and typically have much lower
amplitudes (and hence lower S/N levels in the frequency spectrum) than
coherent pulsations?


 \section{Super-Nyquist spectrum of solar-like oscillations}
 \label{sec:solar}

 \subsection{Impact of finite mode lifetimes}
 \label{sec:life}

The first issue we confront is that solar-like oscillations have
finite lifetimes. In most cases the lifetimes $\tau$ of detectable
oscillations are significantly shorter than the duration $T$ of the
observations (e.g., Dupret et al. 2009). Only for some
gravity-dominated mixed modes of red giants do the lifetimes of
detected oscillations approach or exceed lengths commensurate with
multi-month-long observations. One might therefore not usually expect
to be able to resolve the aforementioned sideband structure in the
frequency domain.

\begin{figure}
 \centerline {\epsfxsize=9.0cm\epsfbox{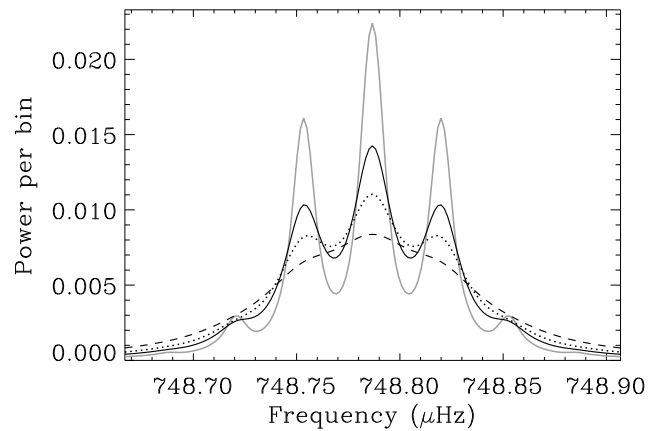}}

 \caption{\small Impact of finite mode lifetimes on the aliased
   quintuplet from the bottom left-hand panel of
   Fig.~\ref{fig:fig2}. The different linestyles show the composite,
   aliased profiles expected when the real mode has a lifetime
   (linewidth) of 73\,days ($0.1\,\rm \mu Hz$; dashed line), 123\,days
   ($0.05\,\rm \mu Hz$; dotted line), 184\,days ($0.02\,\rm \mu Hz$;
   dark solid line) and 1\,yr ($0.01\,\rm \mu Hz$; grey solid line).}

 \label{fig:fig3}
\end{figure}

The underlying noise-free or ``limit-spectrum'' profiles due to
solar-like oscillations are to good approximation Lorentzian in shape,
with the \textsc{fwhm} of the peaks given by $\Gamma = (\pi
\tau)^{-1}$ (e.g., see Chaplin et al. 2002). Fig.~\ref{fig:fig3} shows
how a finite mode lifetime would affect the appearence of the aliased
quintuplet from the bottom left-hand panel of Fig.~\ref{fig:fig2}. The
different linestyles show the composite, aliased limit-spectrum
profiles expected when the real mode has a lifetime (linewidth) of
73\,days ($0.1\,\rm \mu Hz$; dashed line), 123\,days ($0.05\,\rm \mu
Hz$; dotted line), 184\,days ($0.02\,\rm \mu Hz$; dark solid line) and
1\,yr ($0.01\,\rm \mu Hz$; grey solid line).

Lifetimes of solar-like oscillations in main-sequence and subgiant
stars are typically of the order of several days in length. The
lifetimes of detected radial mode oscillations in red giants can be as
long as the 73-day lifetime considered above (e.g., Corsaro et
al. 2012).  The corresponding dashed line in Fig.~\ref{fig:fig3} would
be barely indistinguishable from a Lorentzian profile in the presence
of real noise. The same would of course be true for shorter mode
lifetimes.  The longer lifetimes modelled in Fig.~\ref{fig:fig3} show
much more significant departures from a Lorentzian
appearence. Provided S/N levels in the observed modes are sufficiently
high, it might therefore be possible to discriminate visually the
sidebands of some long-lived modes. It is worth noting once more that
amplitudes of solar-like oscillations are usually much lower than for
coherent pulsators, making it harder to distinguish the sideband
structure.

We conclude that an important consequence of the finite mode lifetimes
is that in most cases one would no longer be able to distinguish
aliased from true oscillation peaks on account of the sideband
structure. However, for solar-like oscillators we are fortunate in
that we do not need to rely on such discrimination alone.


 \subsection{Use of a priori information for solar-like oscillators}
 \label{sec:prior}

\begin{figure*}
 \centerline {\epsfxsize=11.0cm\epsfbox{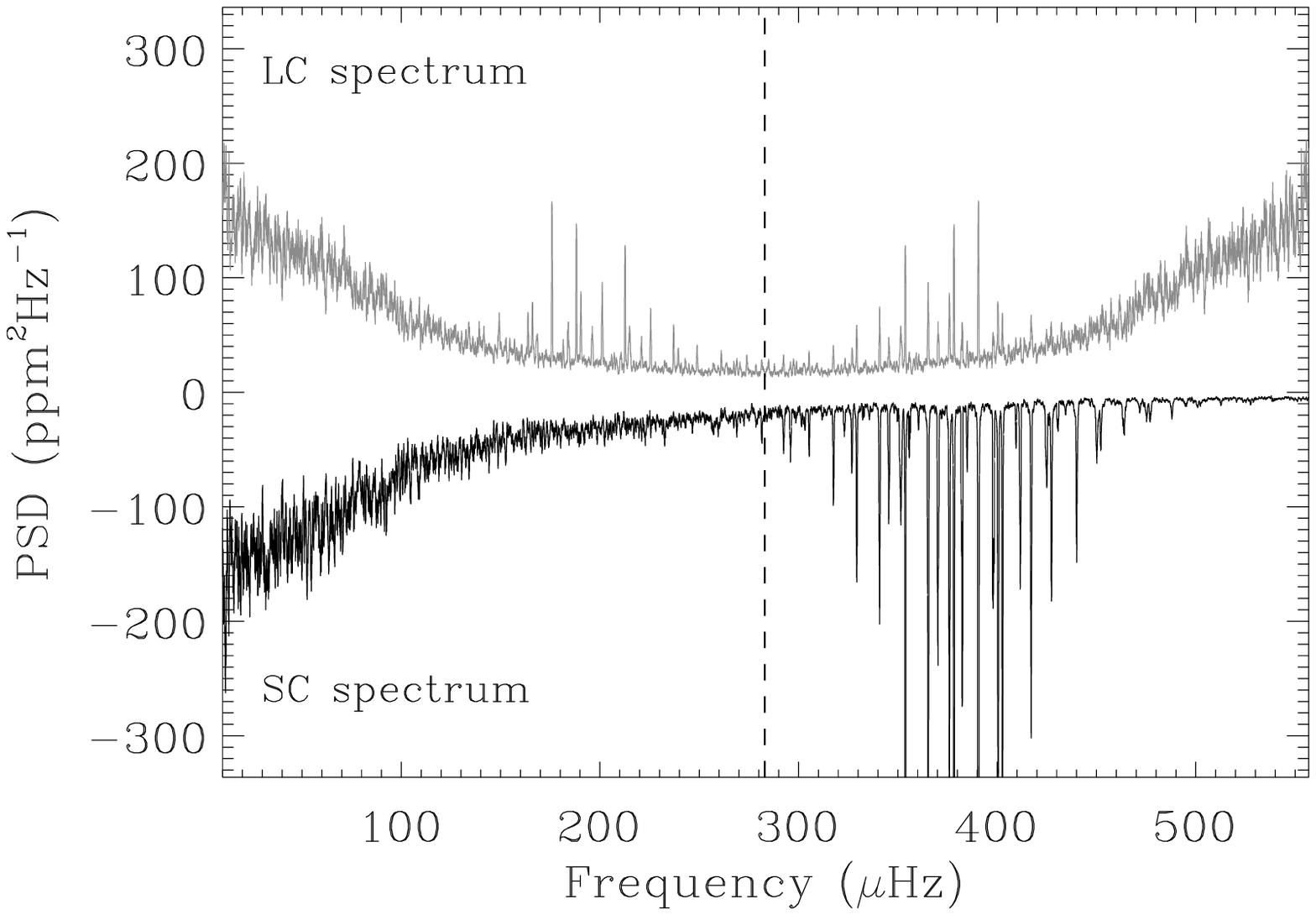}}
 \centerline {\epsfxsize=11.0cm\epsfbox{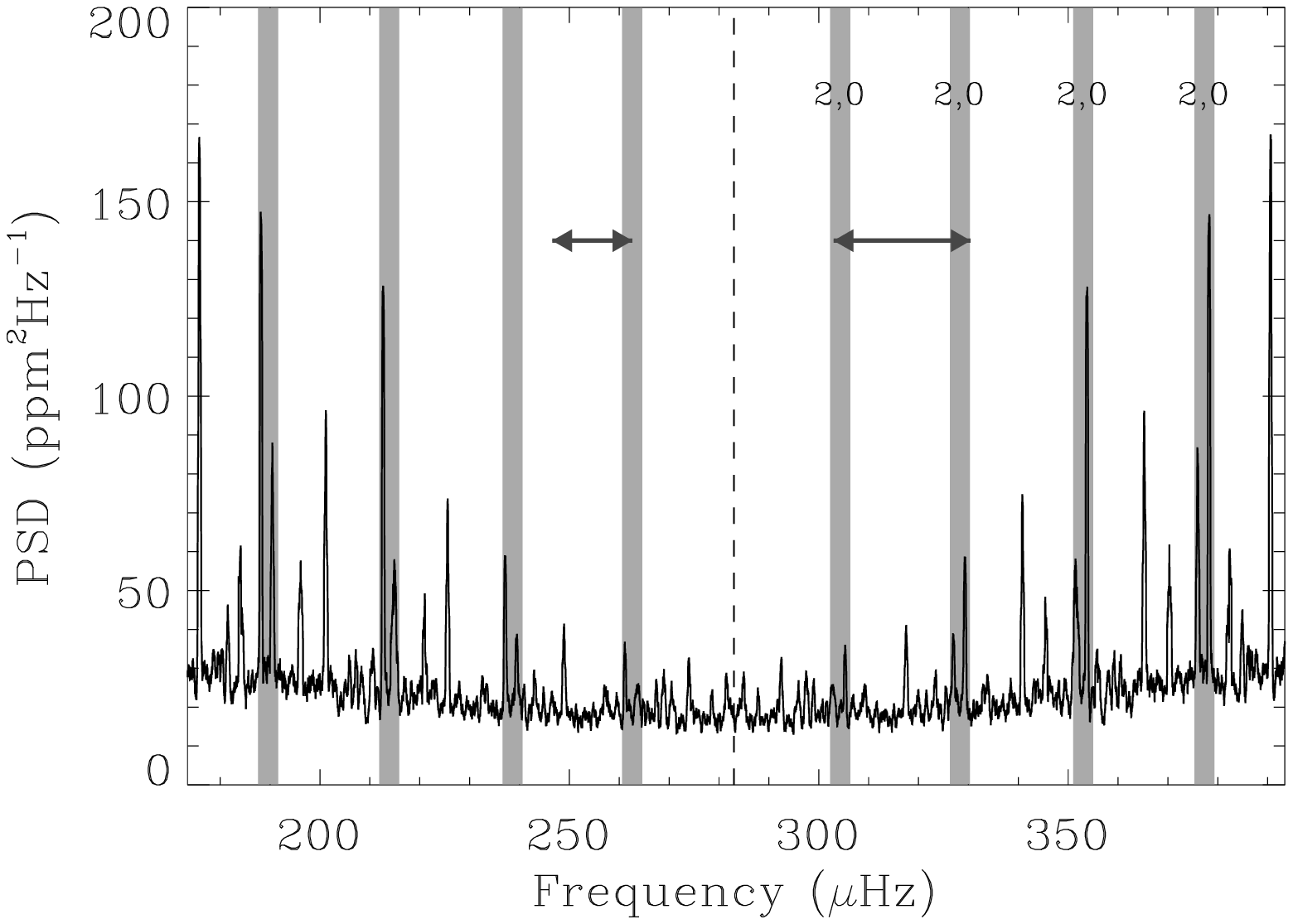}}

 \caption{\small Top panel: frequency power spectra of simultaneous LC
   (upper plot) and SC (lower plot) data collected by \emph{Kepler} on
   the low-luminosity red-giant KIC\,4351319. Bottom panel: zoom of
   the LC spectrum above and below $\nu_{\rm Nyq}$. The lengths of the
   horizontal arrows in the bottom panel correspond to expected
   average separations, assuming a $\nu_{\rm max}$ of $\simeq 190\,\rm
   \mu Hz$ (peaks below $\nu_{\rm Nyq}$) and $\simeq 380\,\rm \mu Hz$
   (peaks above $\nu_{\rm Nyq}$). The shaded regions mark the
   locations of pairs of adjacent $l=2$ and $l=0$ modes.}

 \label{fig:fig4}
\end{figure*}

The rich spectra of overtones shown by solar-like oscillators provides
the necessary, a priori information to allow us to select the true
spectrum from the aliased spectra. We illustrate how with a real
example, the low-luminosity red giant KIC\,4351319. This target was
observed in both long and short cadence during the nominal
\emph{Kepler} Mission. Here, we have used data prepared using the
PDC-MAP pipeline (Stumpe et al. 2012; Smith et al. 2012). The top
panel of Fig.~\ref{fig:fig4} shows the frequency power spectra of
simultaneous LC (upper plot) and SC (lower plot) PDC-MAP data from
\emph{Kepler} Quarters Q8 through Q17. The nominal LC Nyquist
frequency, $\nu_{\rm Nyq}$, is marked by the vertical dashed line. The
luxury afforded by having contemporaneous data available also in SC --
where the oscillations are oversampled -- of course allows us to tag
the peaks in the LC spectrum above $\nu_{\rm Nyq}$ as the true
oscillation frequencies (see the figure). However, two checks based on
use of the LC spectrum alone allow us to make the same call.


The bottom panel of Fig.~\ref{fig:fig4} shows a zoom of the LC
spectrum above and below $\nu_{\rm Nyq}$. Our first check relates to
the large frequency separation, $\Delta\nu$, the spacing shown by
consecutive overtones of the same angular (spherical) degree, $l$. An
extensive body of literature has now established a fairly tight
correlation between the average large separation and the observed
frequency of maximum oscillation power, $\nu_{\rm max}$ (e.g., see
Hekker et al. 2009; Stello et al. 2009; Huber et al. 2011; Mosser et
al. 2012). This relation may be expressed in the form $\Delta\nu
\propto \nu_{\rm max}^{\,\beta}$, with $\beta \simeq 0.76$ for the
cool subgiants and low-luminosity red giants of interest here. While
there is evidently spread in the relation due to, for example,
dependences on mass and metallicity, the strong correlation
nevertheless allows us to discriminate the true from the aliased
spectrum. The lengths of the horizontal arrows in the bottom panel
correspond to expected average separations, assuming a $\nu_{\rm max}$
of $\simeq 190\,\rm \mu Hz$ (peaks below $\nu_{\rm Nyq}$) and $\simeq
380\,\rm \mu Hz$ (peaks above $\nu_{\rm Nyq}$). The shaded regions
mark the locations of pairs of adjacent $l=2$ and $l=0$ modes. The
frequency intervals between these regions provide a visual estimate of
the observed large separation. Evidently, it is the spectrum above
$\nu_{\rm Nyq}$ that conforms with the expected average separation.

The second check relates to the relative power shown by adjacent $l=2$
and $l=0$ modes. Within each pair, the $l=0$ mode lies at higher
frequency. We expect to see more power in the $l=0$ mode than in its
$l=2$ counterpart. Whilst predictions of the exact power ratios are
rendered uncertain by the complexities of non-adibatic calculations,
observed ratios are usually not too far from the $\simeq 2$-to-1 ratio
expected from the assumption of energy equipartition and geometric
cancellation of the perturbations on the visible stellar disc (when
projected on to spherical harmonic functions, with an appropriate
limb-darkening law; e.g., see Aerts et al. 2009). Unless the star is
observed with the rotation axis along the line-of-sight, power shown
by the $l=2$ modes will be spread across several components (e.g.,
Gizon \& Solanki 2003). The ratio in total power then tends to be
exaggerated in the frequency power spectrum because it is the
\emph{maximum} power spectral densities, i.e., the heights of the mode
peaks, that are immediately apparent from a visual inspection.

Inspection of the mode pairs in the bottom panel of
Fig.~\ref{fig:fig3} again implies that the spectrum above $\nu_{\rm
  nyq}$ is the true one: The $l=0$ modes, the higher-frequency modes
in each marked pair, have the higher observed powers.

 \subsection{Impact of signal attenuation and background aliasing}
 \label{sec:sinc}

How far above $\nu_{\rm Nyq}$ might we hope to detect solar-like
oscillations in the LC \emph{Kepler} data? A combination of several
factors means that pushing the limit well above $\nu_{\rm Nyq}$ is
challenging.  First, the sinc-function attenuation due to the finite
integration time leads to significant reduction of the observed
oscillation amplitudes at frequencies above $\nu_{\rm Nyq}$. Second,
the impact of the sinc attenuation is exacerbated by the fact that the
intrinsic maximum amplitudes of solar-like oscillations decrease with
increasing $\nu_{\rm max}$.  And third, we must also contend with
background power aliased from the region below $\nu_{\rm Nyq}$, which
has contributions from stellar granulation, stellar activity, shot and
instrumental noise. Note that any non-white or ``red'' noise
component, such as the granulation, will also have a contribution above
$\nu_{\rm Nyq}$ that is sinc-function attenuated, just like the
oscillations (e.g., see Kallinger et al. 2014).

We come back to the example of KIC\,4351319 to help illustrate these
points. The top panel of Fig.~\ref{fig:fig4} shows clearly the
attenuated power of the oscillations in the LC spectrum of
KIC\,4351319, relative to those observed in the SC spectrum. The
impact of the sinc-function attenuation on the observed S/N is
compounded in the LC spectrum because the background in the region of
$\nu_{\rm max}$ is the aliased background from lower frequencies,
which is dominated by granulation \emph{and} is much higher -- by more
than an order of magnitude -- than the shot-noise background.  The
spectrum of the oscillations is in reality coincident in frequency
with the lower-power, higher-frequency part of the granulation
spectrum, as shown by the SC spectrum of KIC\,4351319.  A much higher
S/N is therefore observed in the oscillations when using SC data,
because we do not have to contend with aliasing of power from the
lower-frequency, higher-power part of the granulation spectrum. Our
tests indicate that it is only when $K_{\rm p}$ is fainter than
approximately 11 to 12\,mag that the photon shot noise becomes an
important factor in determining the detectability of the modes in the
LC super-Nyquist regime.


\begin{figure*}
 \centerline {\epsfxsize=11.0cm\epsfbox{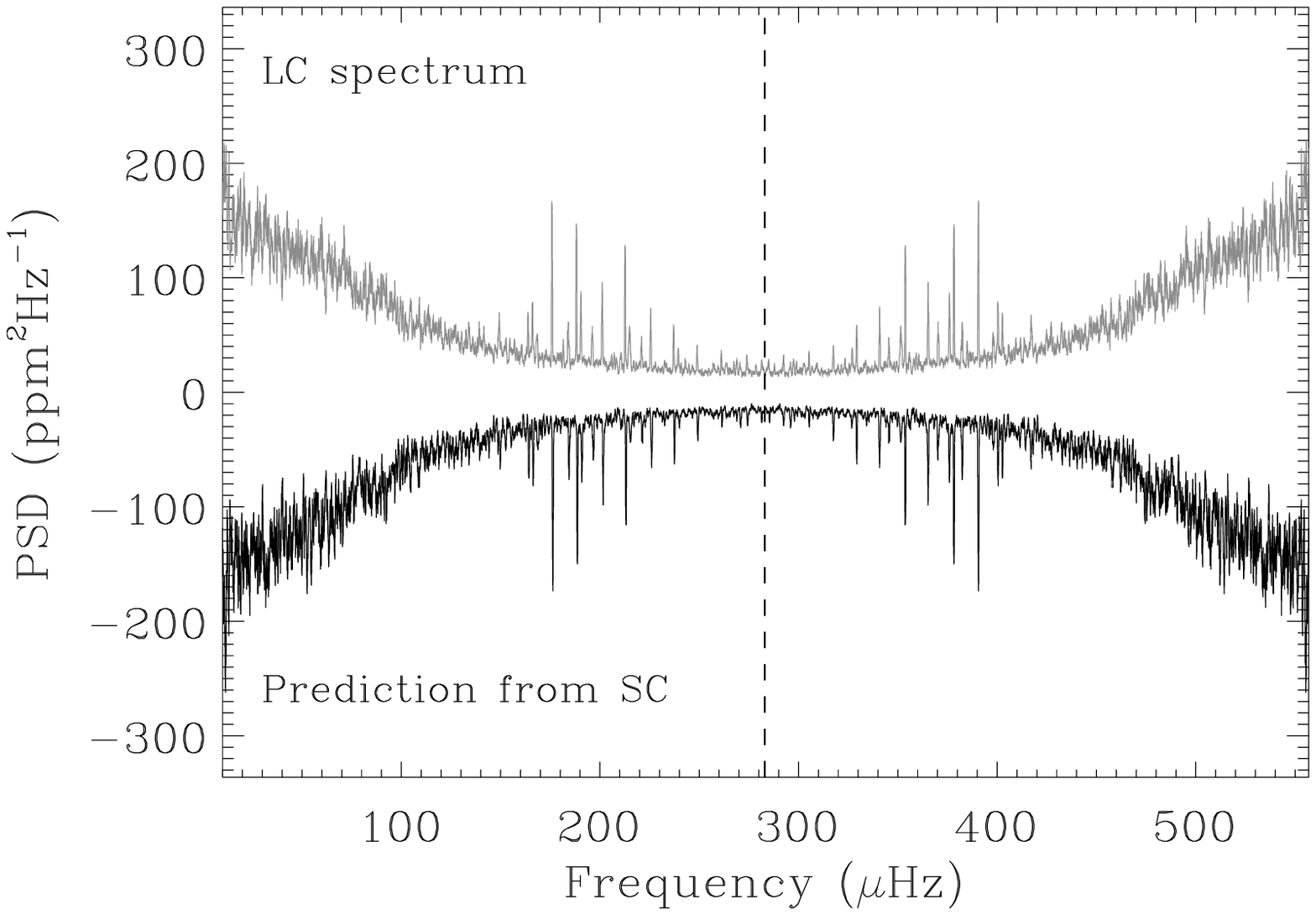}}

 \caption{\small Frequency power-spectrum of KIC\,4351319: Upper plot
   shows LC spectrum, lower plot predicted LC spectrum made from the
   available SC spectrum.}

 \label{fig:fig5}
\end{figure*}


From the observed SC spectrum of KIC\,4351319 we may construct an
(almost) accurate prediction of the LC spectrum. We begin by
multiplying the observed SC spectrum\footnote{The signatures of the
  oscillations and granulation are also attenuated in the SC data, but
  since the SC Nyquist frequency is $\simeq 8496\,\rm \mu Hz$ the
  impact in the frequency range of interest here is negligible.} by
the sinc-squared attenuation filter, $\eta^2$. (Note we have not
bothered to attempt to discriminate the white contribution due to shot
noise, since it is very small compared to the other contributions.)
The predicted LC power below $\nu_{\rm Nyq}$ is given by adding the
sub-Nyquist SC attenuated power to the aliased, super-Nyquist
attenuated spectrum from above $\nu_{\rm Nyq}$ (after reflecting the
latter about $\nu_{\rm Nyq}$). The predicted LC power above $\nu_{\rm
  Nyq}$ is then just the reflected prediction from below $\nu_{\rm
  Nyq}$. Fig.~\ref{fig:fig5} shows the resulting prediction (lower
plot), together with the LC spectrum from Fig.~\ref{fig:fig4} (upper
plot).


\begin{figure*}
 \centerline {\epsfxsize=9.0cm\epsfbox{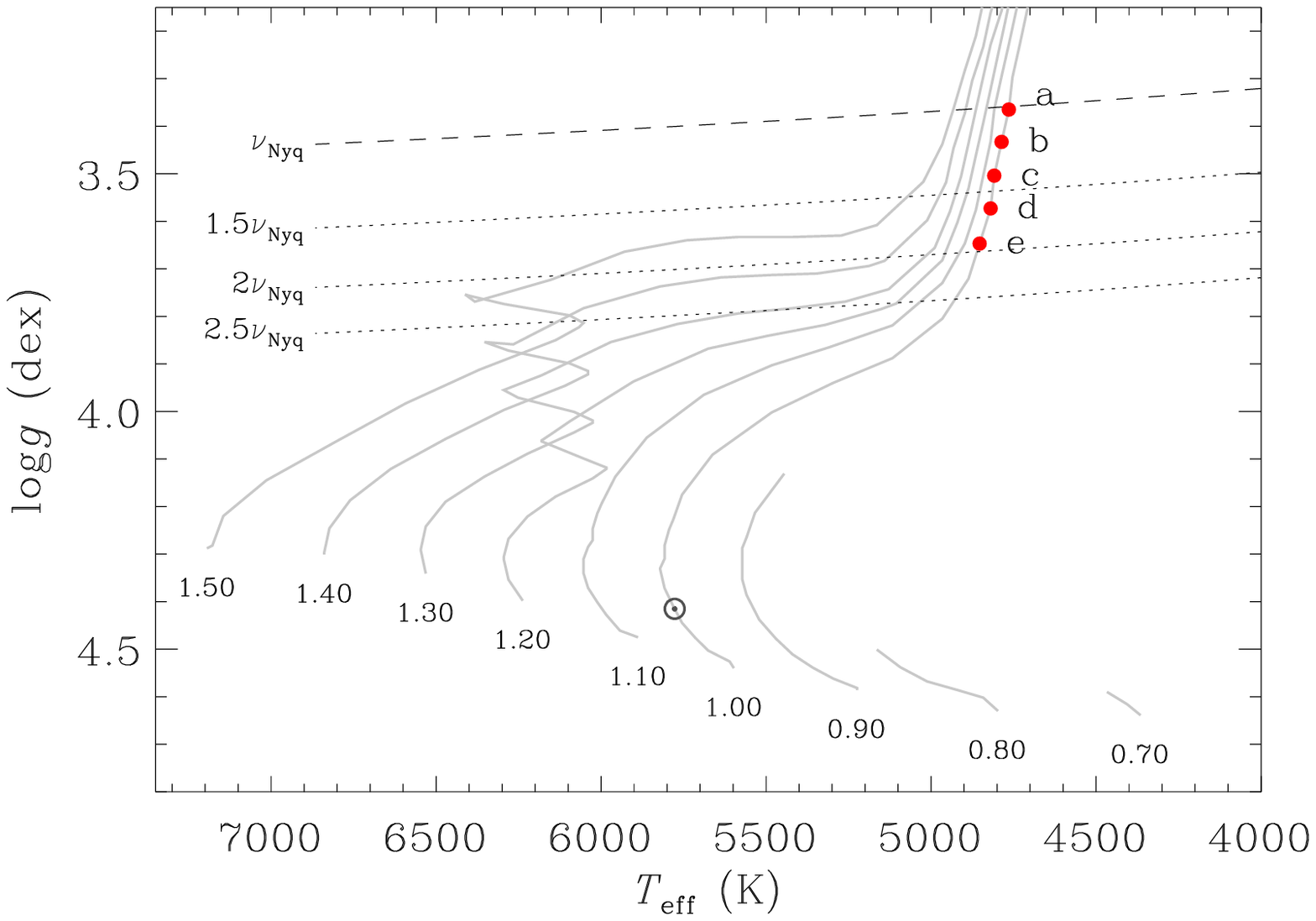}
              \epsfxsize=9.0cm\epsfbox{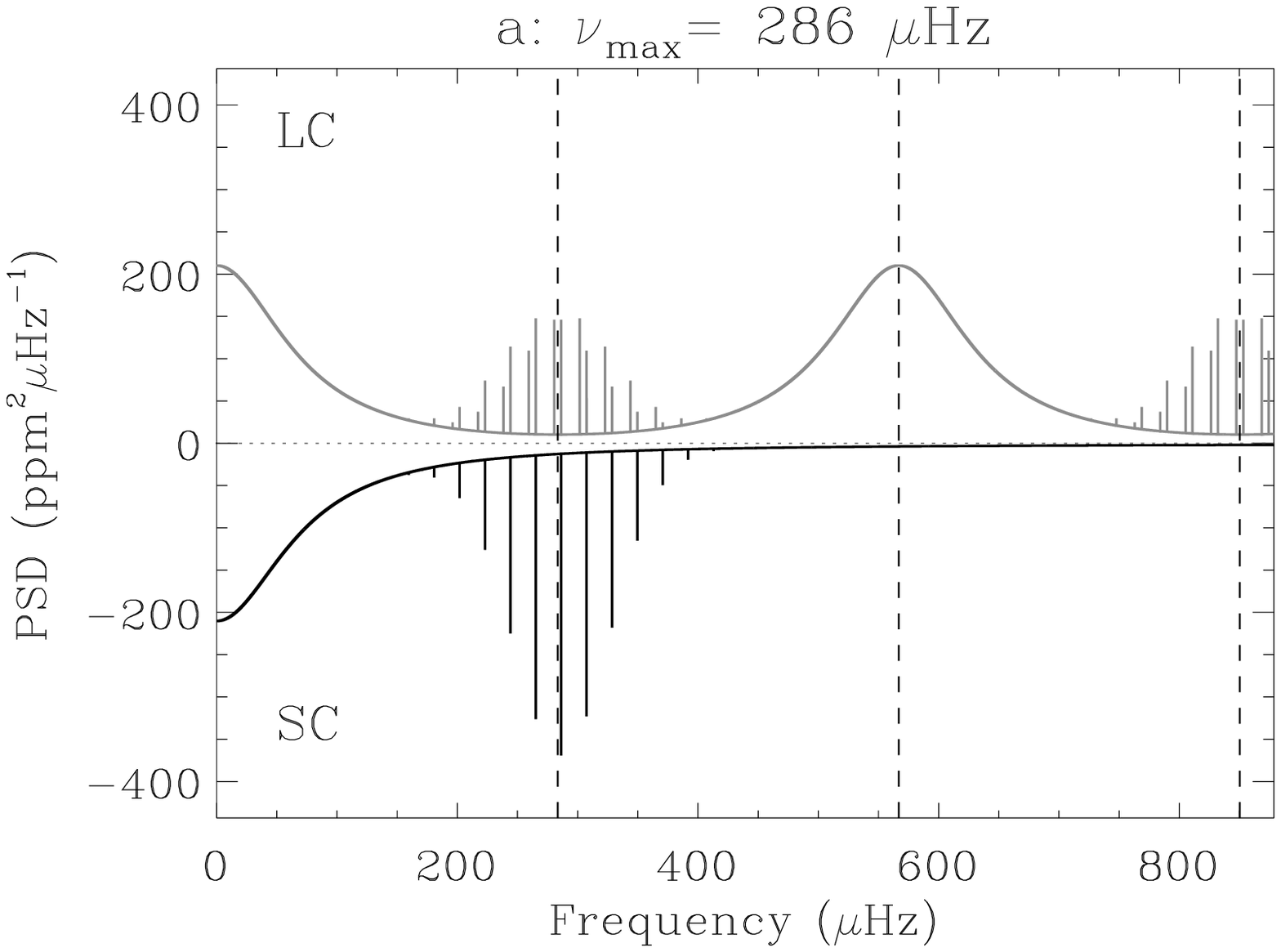}}
 \centerline {\epsfxsize=9.0cm\epsfbox{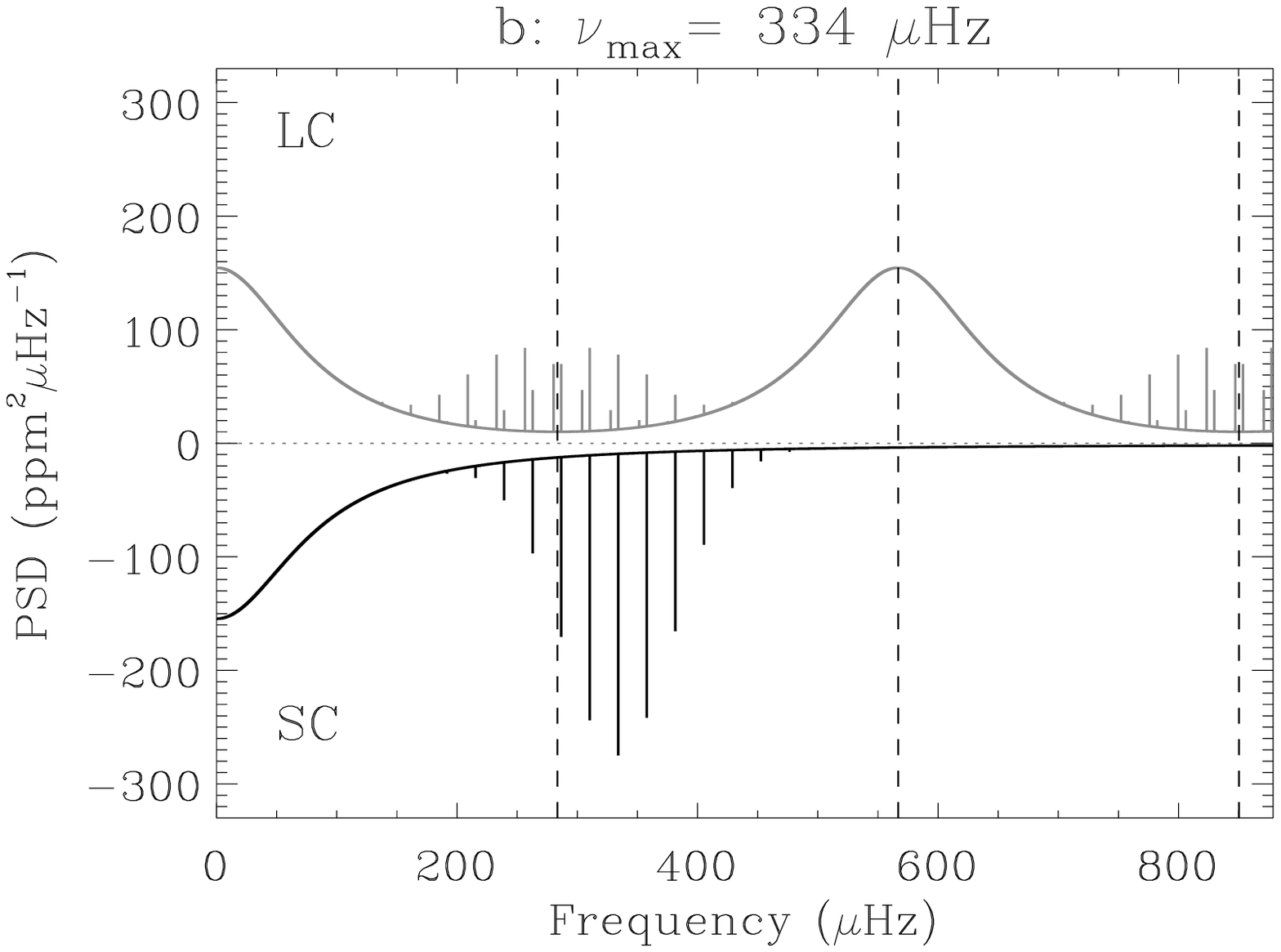}
              \epsfxsize=9.0cm\epsfbox{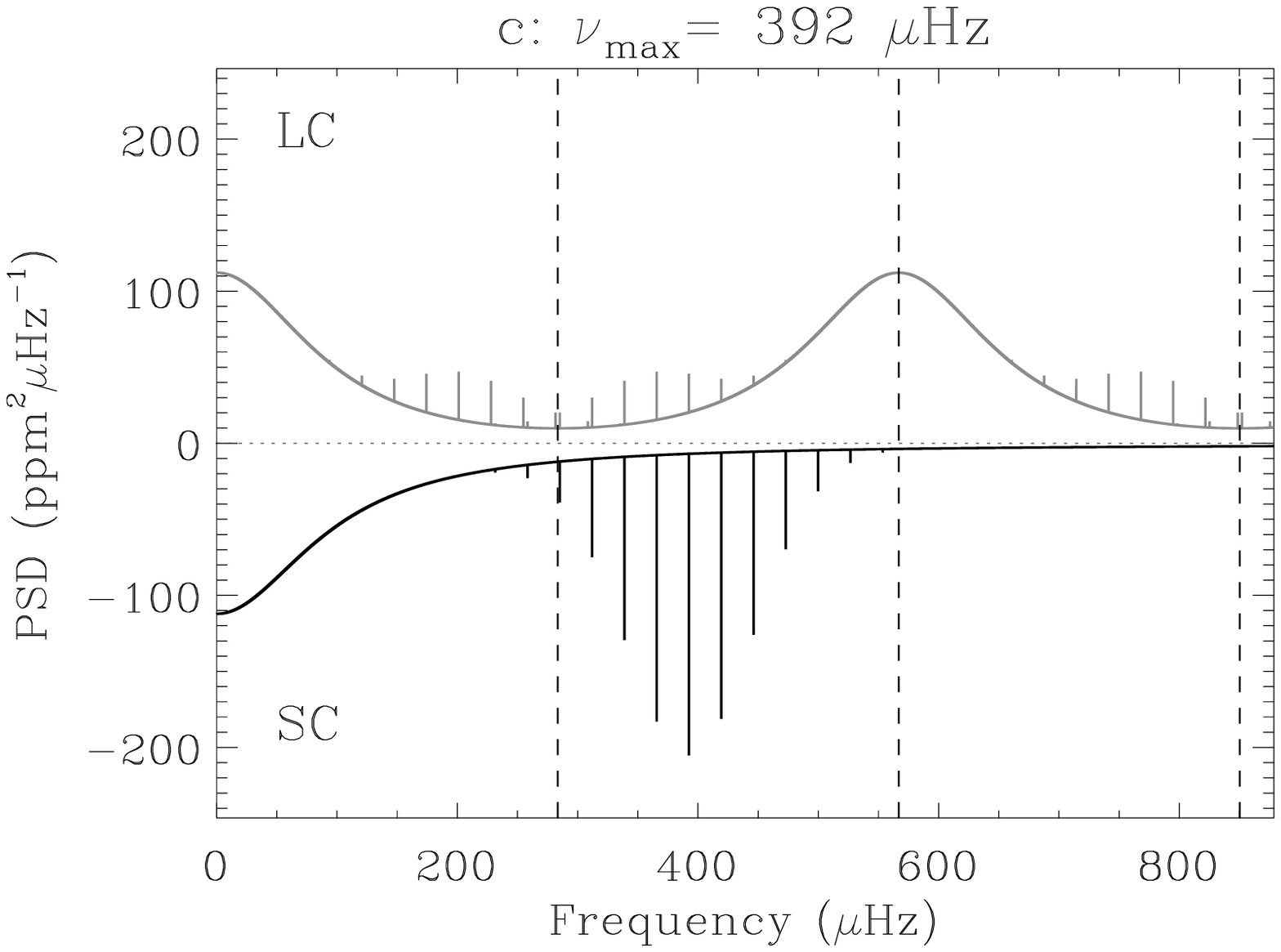}}
 \centerline {\epsfxsize=9.0cm\epsfbox{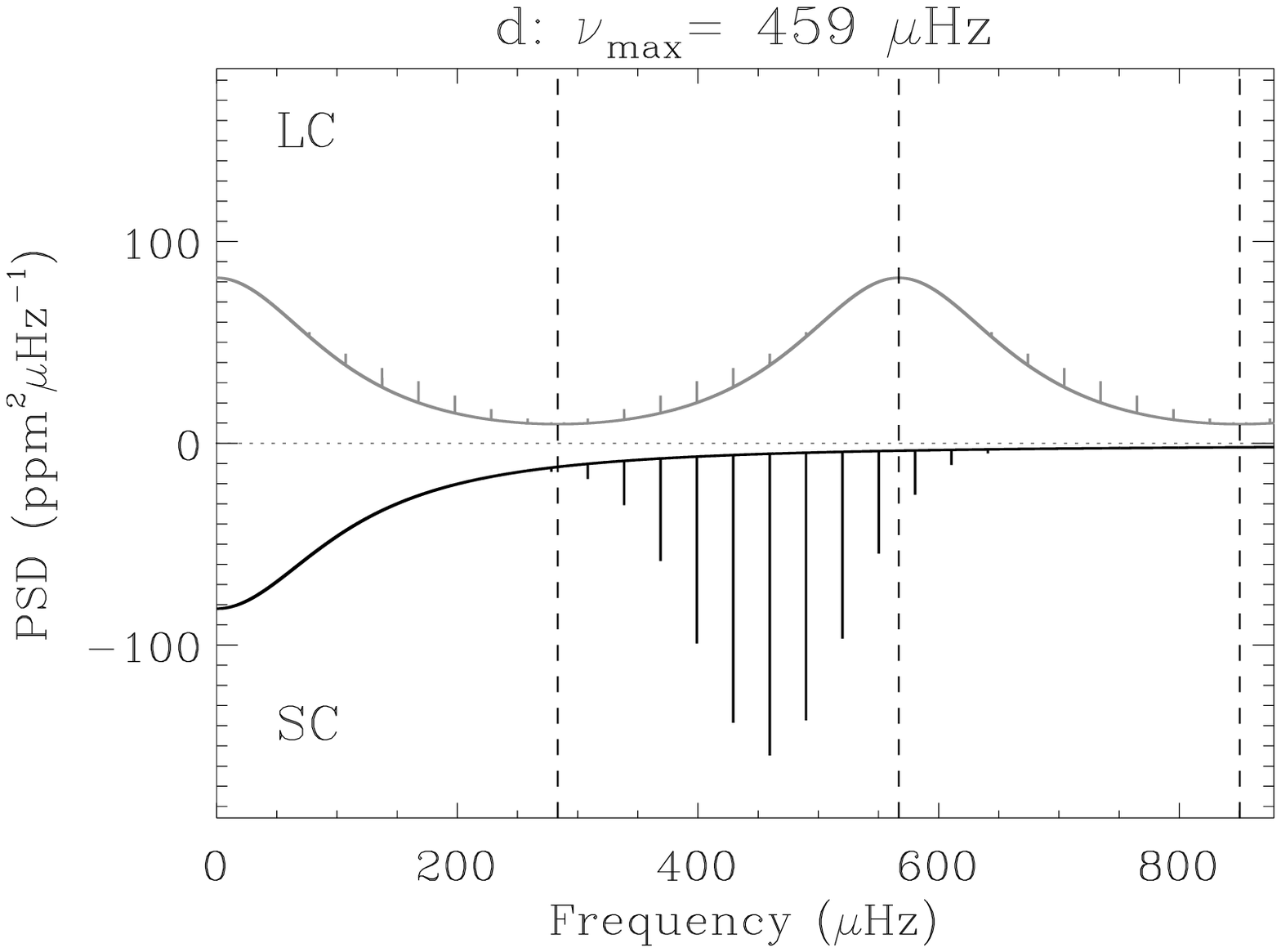}
              \epsfxsize=9.0cm\epsfbox{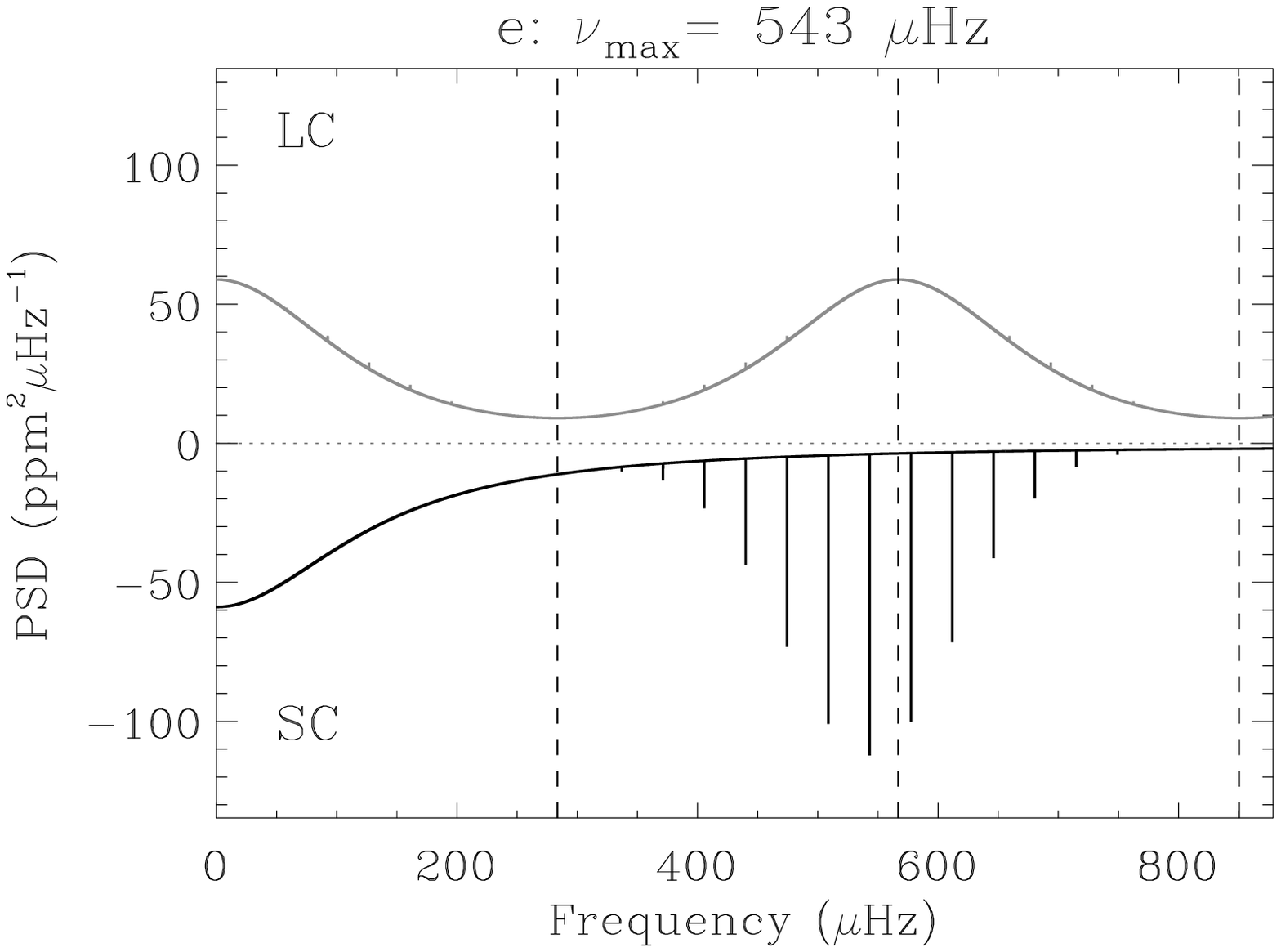}}

 \caption{\small Idealized guide to the changing appearence of
   undersampled and oversampled oscillations spectra with varying
   $\nu_{\rm max}$, based on predictions using scaling relations
   applied to the fundamental properties of several $1\,\rm M_{\odot}$
   stellar models. The top left-hand panel shows the locations of the
   models in the $\log g$-$T_{\rm eff}$ plane. The other panels show
   schematic representations of the limit (noise-free) power density
   spectra expected for LC (undersampled, upper plots) and SC
   (oversampled, lower plots) data, assuming each model was observed
   as a bright \emph{Kepler} target having a \emph{Kepler} apparent
   magnitude of $K_{\rm p} = 9$.  Spectra are plotted up to a
   frequency of just over $3\nu_{\rm Nyq}$ -- with dashed lines
   marking $\nu_{\rm Nyq}$ boundaries -- to fully include the SC
   oscillation spectrum of the highest $\nu_{\rm max}$ (least evolved)
   stellar model (see plot titles).}

 \label{fig:fig6}
\end{figure*}


When the oscillations are undersampled, the appearence of the
oscillations spectra in the super-Nyquist regime depends critically on
the proximity of $\nu_{\rm max}$ to the boundary at $2\nu_{\rm Nyq}$.
This boundary is where the first zero of the sinc-function attenuation
lies, and is also where aliased background power -- from granulation,
activity and instrumental noise -- is most severe, coming as it does
from the very low-frequency part of the spectrum.  Fig.~\ref{fig:fig6}
is an idealized, visual guide to the changing appearence of spectra as
$\nu_{\rm max}$ is varied. The figure shows schematic frequency-power
spectra made by applying suitable scaling relations (see below) to the
properties of a sequence of $1\,\rm M_{\odot}$ stellar evolutionary
models computed by the Padova group (Marigo et al. 2008). The top
left-hand panel of Fig.~\ref{fig:fig6} plots tracks in the $\log
g$-$T_{\rm eff}$ plane, with the selected models shown in red. All
five models have a predicted frequency of maximum solar-like
oscillations power, $\nu_{\rm max}$, that lies above $\nu_{\rm
  Nyq}$. The characteristic $\nu_{\rm max}$ frequency was estimated
from the fundamental stellar properties (Brown et al. 1991; Kjeldsen
\& Bedding 1995) using:
 \begin{equation}
 \nu_{\rm max} = \nu_{\rm max,\odot}\, \left( \frac{M}{\rm M_{\odot}}
 \right) \left( \frac{R}{\rm R_{\odot}} \right)^{-2} \left(
 \frac{T_{\rm eff}}{{\rm T_{eff,\odot}}}\right)^{-0.5},
 \label{eq:numax}
 \end{equation}
where we have scaled against the solar values of $\nu_{\rm max,\odot}
= 3090\,\rm \mu Hz$ and ${\rm T_{eff,\odot}} = 5777\,\rm K$.  Loci of
constant $\nu_{\rm Nyq}$, and selected multiples thereof, are also
marked in the top panel. 

After the top left-hand panel of Fig.~\ref{fig:fig6}, the other panels
show schematic representations of the limit (noise-free) power density
spectra expected for LC (upper plots) and SC (lower plots) data, where
we assumed each model was observed as a bright \emph{Kepler} target
having a \emph{Kepler} apparent magnitude of $K_{\rm p} = 9$. The
composite SC spectra were constructed using the scaling relations and
formulae in Chaplin et al. (2011b), with contributions from
oscillations, granulation and shot noise included. We refer the reader
to that paper for further details.

There was one ingredient missing from Chaplin et al. (2011b) that we
needed to make the spectra, and that was the step to convert estimates
of the maximum mode amplitudes, $A_{\rm max}$, into maxmium mode power
spectral densities (peak heights) $H_{\rm max}$. In this paper we have
used an appropriate formalism from Chaplin et al. (2008), i.e.,
 \begin{equation}
 H_{\rm max} = \frac{2A_{\rm max}^2}{\pi T \Gamma + 2},
 \end{equation}
with mode peak linewidths $\Gamma$ estimated from Appourchaux et
al. (2012) using the effective temperature $T_{\rm eff}$ of the
models.  To avoid cluttering the plots we only mark the predicted
limit (noise-free) heights of the radial ($l=0$) modes; we do not plot
the Lorentzian limit profiles or predictions for non-radial modes. The
predicted oscillation lifetimes (Lorentizian peak linewidths) are in
all cases short (large) enough that the timing sidebands would be
unresolved (see Section~\ref{sec:life}), and so each range of
$\nu_{\rm Nyq}$ would essentially be a near identical, reflected copy.

The predicted SC spectra in Fig.~\ref{fig:fig6} show where the true
frequencies lie. Regarding the general appearence of the SC spectra:
The power in, and the characteristic timescale of, the granulation
both increase with decreasing $\nu_{\rm max}$; we also recall that
the power in the oscillations increases with decreasing $\nu_{\rm
  max}$.  The nature of these changes is such that all composite SC
spectra are, in their gross properties at least, approximately
homologously scaled versions of one another.

The SC spectra were converted to predicted LC spectra by applying the
sinc-squared attenuation in power, followed by suitable aliasing about
the $\nu_{\rm Nyq}$ boundaries (see Section~\ref{sec:prior}).  The
schematic spectra have been plotted up to a frequency of just over
$3\nu_{\rm Nyq}$, to fully encompass the SC oscillation spectrum of
the highest $\nu_{\rm max}$ (least evolved) stellar model. The
vertical dashed lines mark the $\nu_{\rm Nyq}$ boundaries.  Frequency
aliasing adds to the complexity of the LC spectra, more severely in
some cases than others depending on the placement of the true
frequencies with respect to the multiple $\nu_{\rm Nyq}$ boundaries.
These plots show very clearly how attenuation of the oscillations can
have a drastic impact on the observed S/N when the true frequencies
are undersampled and hence lie in the super-Nyquist regime. The
oscillation peaks are barely observable in the highest $\nu_{\rm max}$
case shown here.

Before summarizing results from our simulations, it is worth noting
that we have not included any low-frequency contributions due to
stellar activity or instrumental or data-reduction noise. The
predicted S/N ratios implied by the modelled spectra should therefore
be regarded as upper-limit estimates. Nevertheless, starting from the
detection recipe in Chaplin et al. (2011b), we have estimated
detection probabilities for each of these modelled cases. We folded in
predicted power from the non-radial modes, and modified the detection
algorithm to operate on the composite, aliased spectrum in a single
range of $\nu_{\rm Nyq}$. Our predictions suggest that for the targets
in the nominal-Mission \emph{Kepler} archive with 4\,yr of data, the
upper limit for detections will be around $\nu_{\rm max} \simeq
500\,\rm \mu Hz$. Instrumental noise levels will be higher for K2
(Howell et al. 2014), however, with reference to the above, intrinsic
stellar noise should remain the limiting factor for brighter
targets. Taking $T=75\,\rm days$, the length of each K2 campaign, we
estimate an absolute upper detection limit between $\nu_{\rm max}
\simeq 400$ and $\simeq 450\,\rm \mu Hz$.

 \subsection{Some real examples}
 \label{sec:real}


\begin{figure*}
 \centerline {\epsfxsize=9.0cm\epsfbox{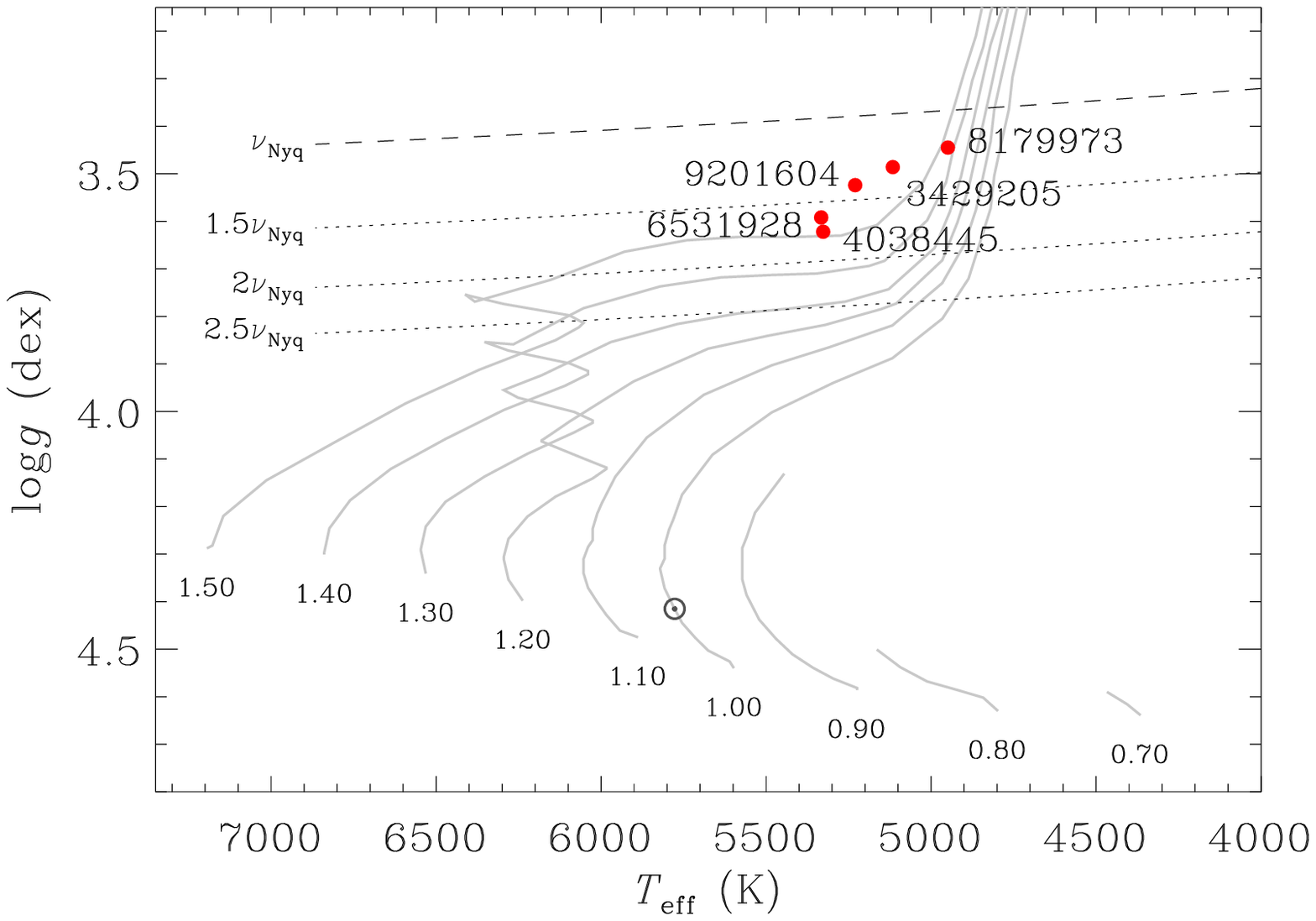}
              \epsfxsize=9.0cm\epsfbox{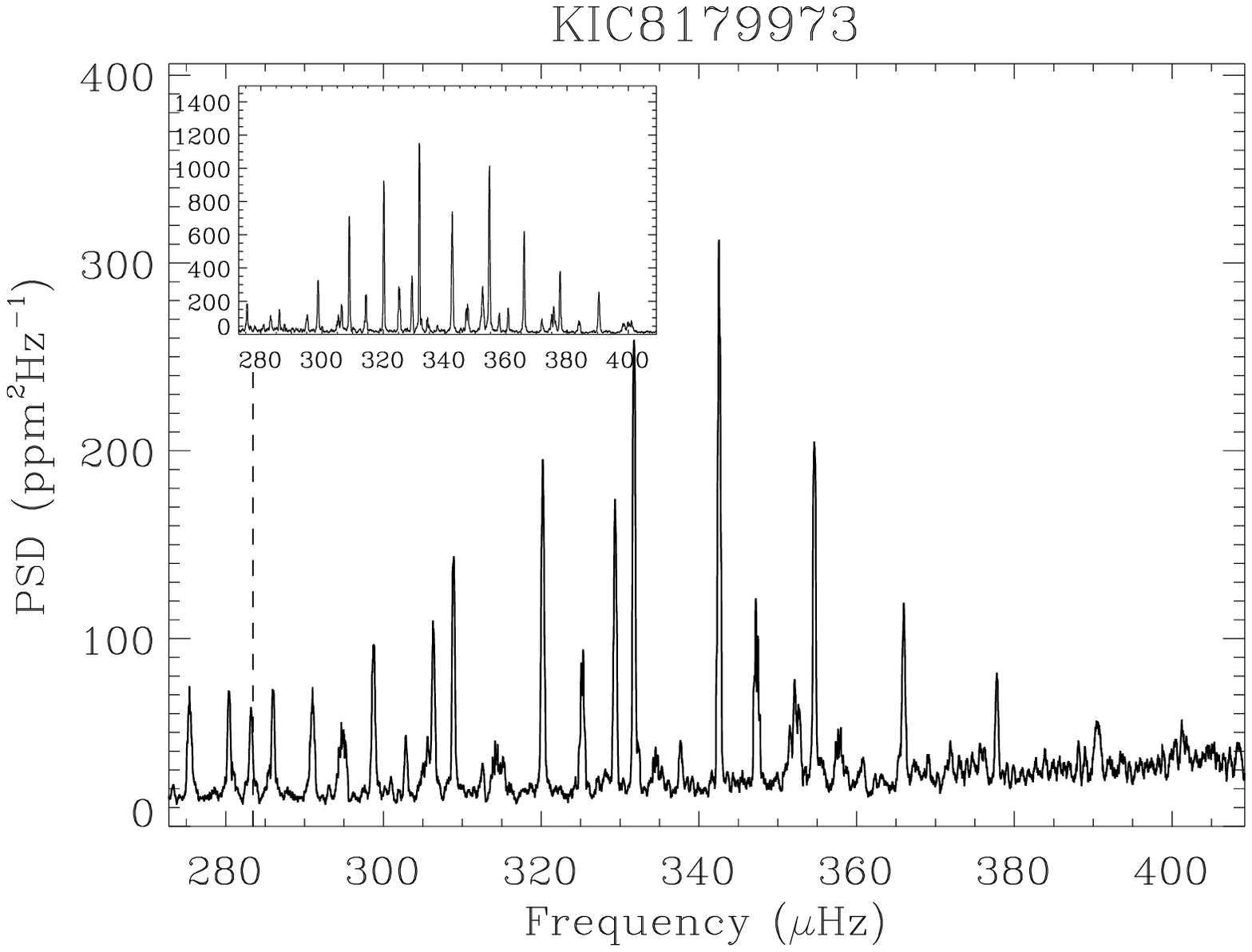}}
 \centerline {\epsfxsize=9.0cm\epsfbox{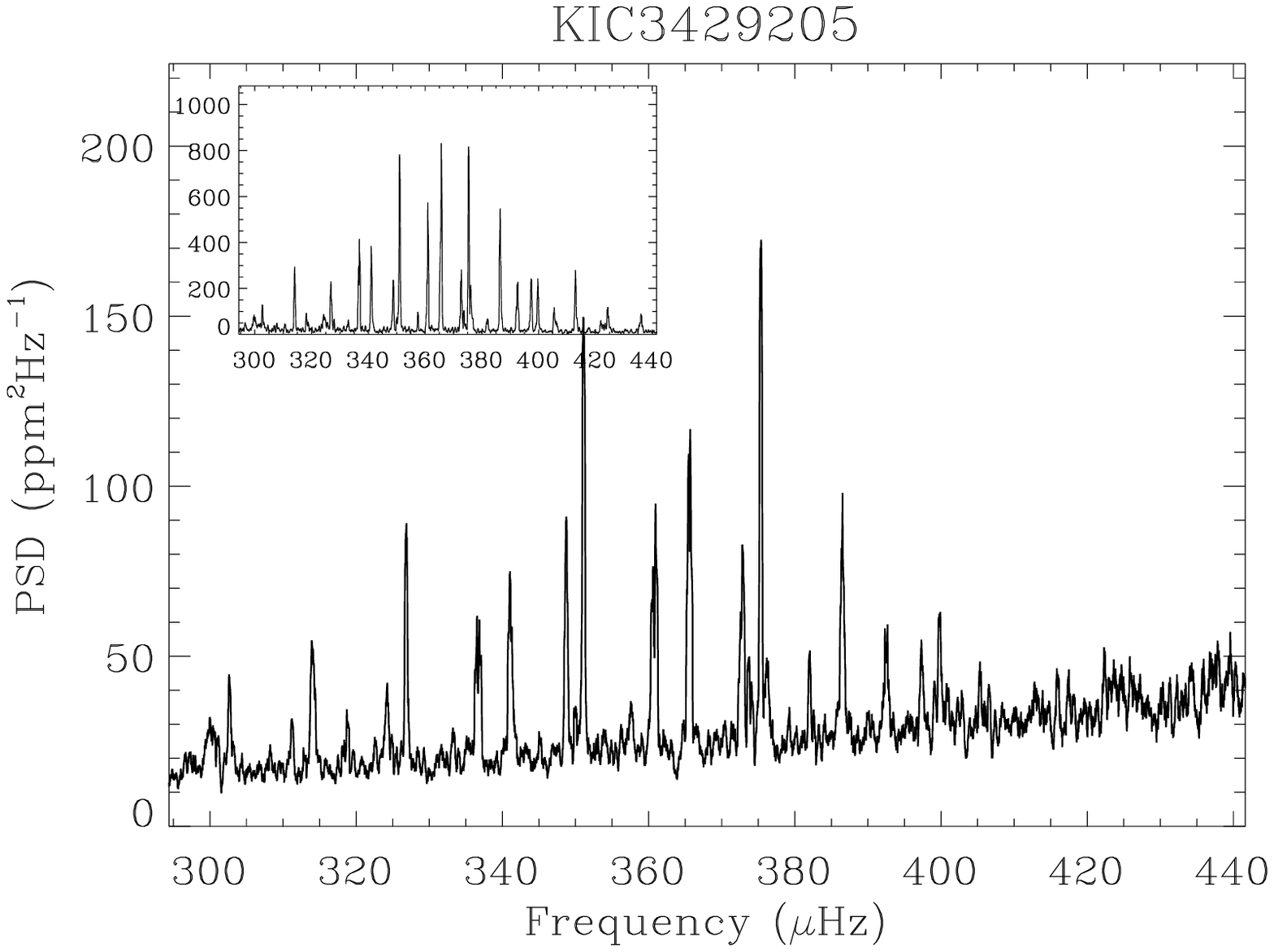}
              \epsfxsize=9.0cm\epsfbox{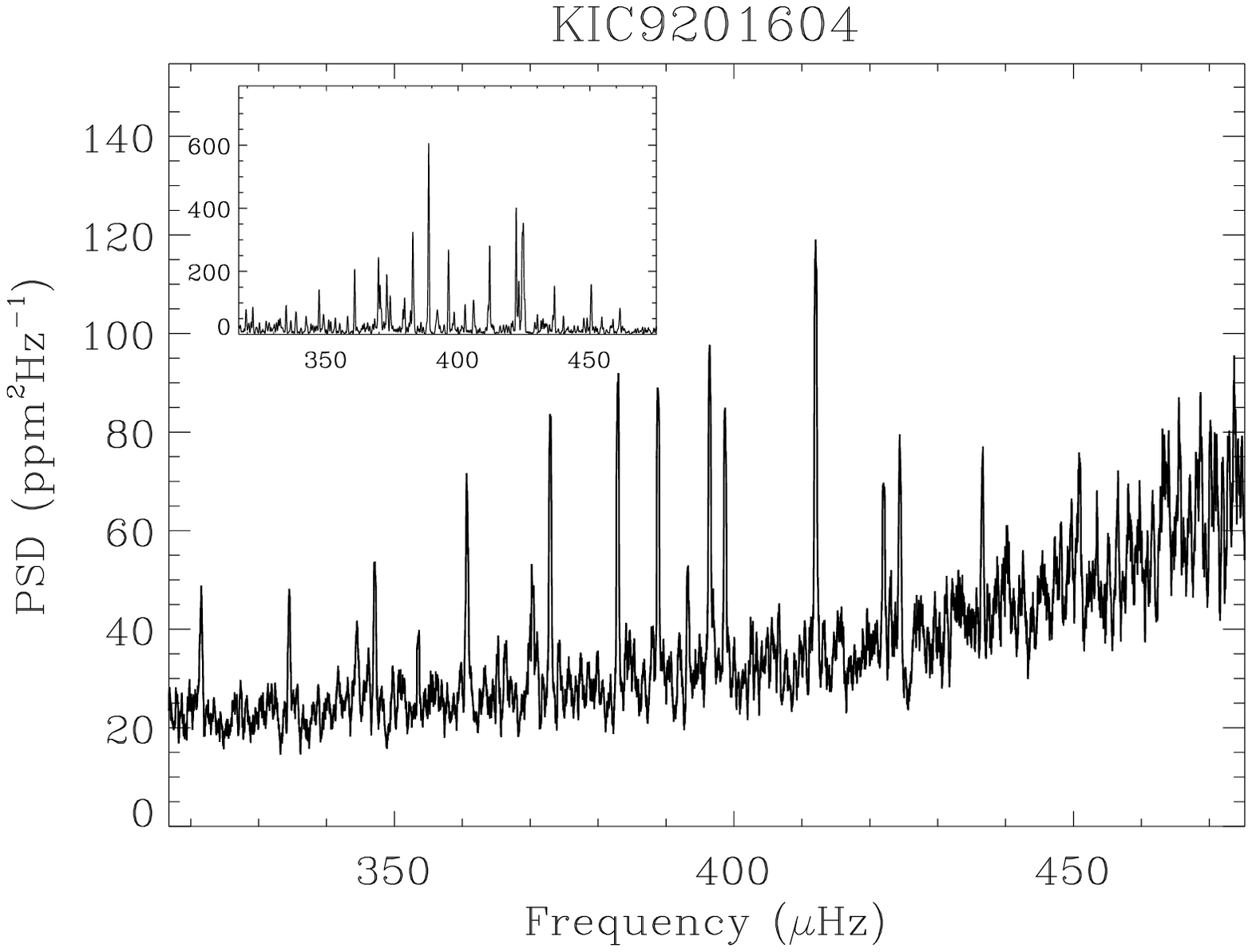}}
 \centerline {\epsfxsize=9.0cm\epsfbox{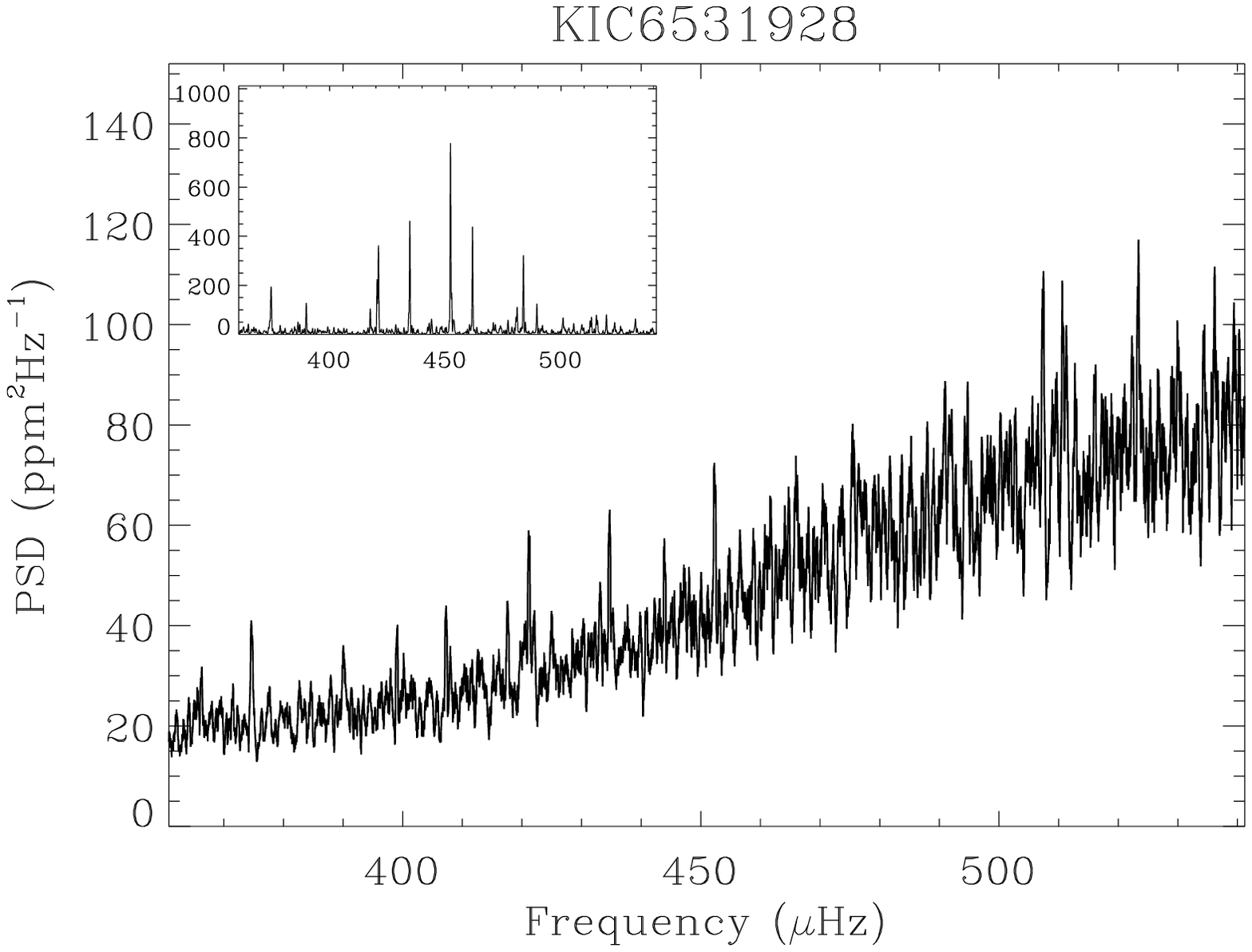}
              \epsfxsize=9.0cm\epsfbox{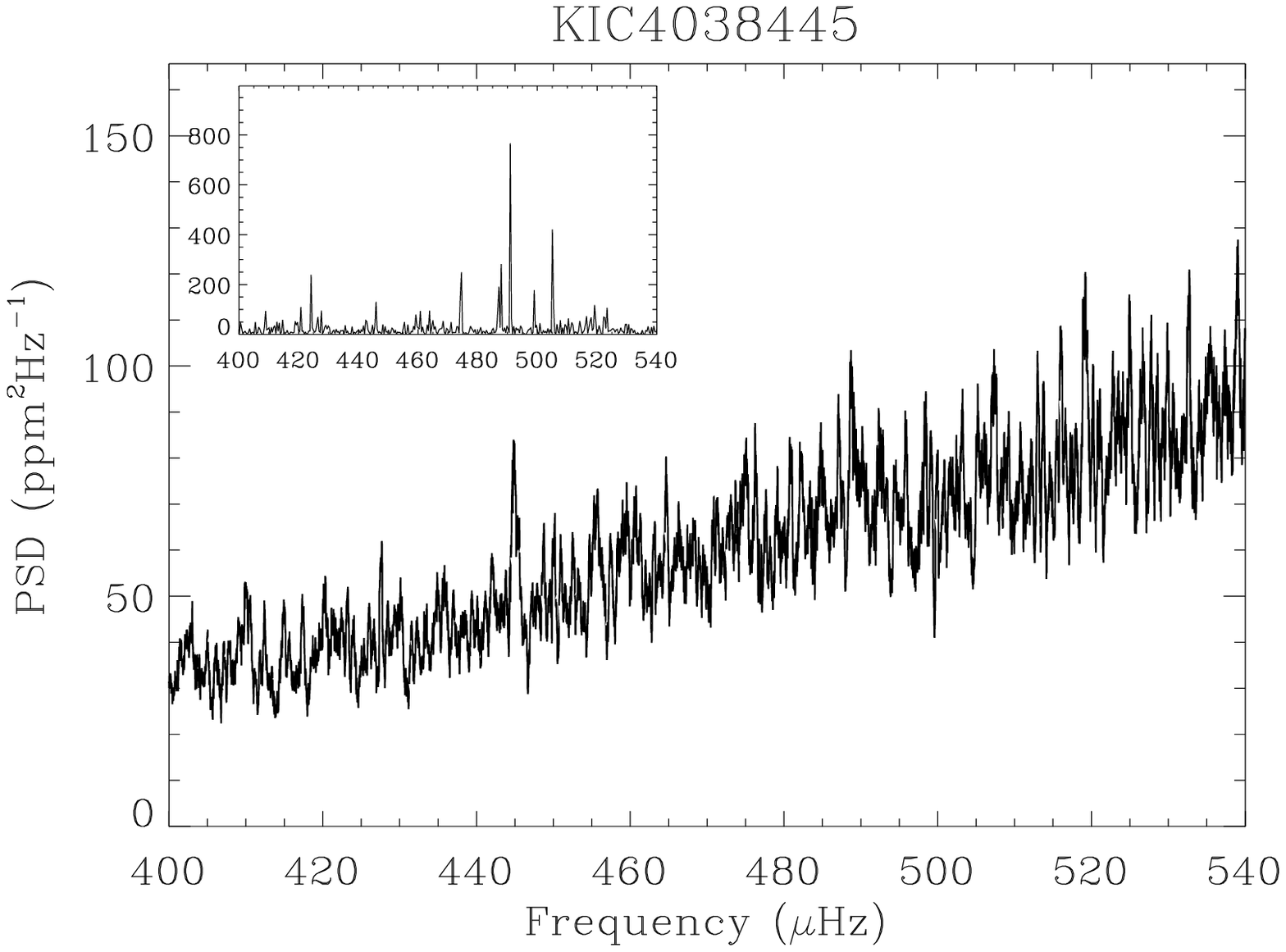}}

 \caption{\small Demonstration, using real \emph{Kepler} LC data on
   five stars, of undersampled solar-like oscillation spectra in the
   LC super-Nyquist regime. Each target also has some SC data
   available, in which the oscillations are oversampled. The top
   left-hand panel shows the locations of the stars in the $\log
   g$-$T_{\rm eff}$ plane. Other panels: main plots show the LC power
   spectra (oscillations undersampled), insets show power spectra
   computed from available SC data (oscillations oversampled).}

 \label{fig:fig7}
\end{figure*}


Fig.~\ref{fig:fig7} shows some more super-Nyquist examples from the
\emph{Kepler} archive. Five targets with \emph{Kepler} LC and SC data
have been selected that have undersampled oscillations in LC. The
$\nu_{\rm max}$ values range from $\simeq 320\,\rm \mu Hz$ (just above
$\nu_{\rm Nyq}$) to $\simeq 550\,\rm \mu Hz$ (just below $2\nu_{\rm
  Nyq}$).  It is worth stressing that there are not many targets like
this in the \emph{Kepler} archive with SC data (in particular at the
low end of the range).  The top left-hand panel plots the locations of
the stars in the $\log g$-$T_{\rm eff}$ plane, with stellar properties
taken from Chaplin et al. (2014).  The main plots of the other panels
show zooms of the LC power spectra in the undersampled, super-Nyquist
regions where the real oscillation frequencies lie. Note that we chose
a slightly different layout compared to the previous figures to show
more clearly the lower S/N cases at higher $\nu_{\rm max}$. The insets
show power spectra computed from available SC data. Note that, unlike
KIC\,4351319, the SC coverage was much more limited for some of these
stars.  Even though the SC and LC data are not necessarily
contemporaneous, the insets show clearly where the real oscillations
are.

From our model predictions above, we expect to have good sensitivity
up to $\nu_{\rm max} \simeq 400\,\rm \mu Hz$, a conclusion that is
borne out by these real data. At frequencies above this, detecting the
modes becomes more difficult.  Signatures of the oscillations are
clearly apparent in the LC spectrum of KIC\,6531928 ($\nu_{\rm max}
\simeq 450\,\rm \mu Hz$) but at a much reduced S/N compared to the
other targets. KIC\,8038445 has a $\nu_{\rm max}$ of $490\,\rm \mu
Hz$. Signatures of a few modes are just detectable in its LC
spectrum. This case marks the approximate upper $\nu_{\rm max}$ limit
for making detections in the archival \emph{Kepler} data.

Things become much more challenging as we move to higher frequencies,
to the region where $\nu_{\rm max}$ straddles or lies close to the
boundary at $2\nu_{\rm Nyq}$. Imposing a high-pass filter to suppress
the very-low frequency background power in the sub-Nyquist regime does
not help to improve the S/N in the super-Nyquist regime: the filtering
response is the same in each domain of $\nu_{\rm Nyq}$ and so any
oscillation signal is also filtered.

 \section{Conclusion}
 \label{sec:conc}

We have considered the prospects for detecting solar-like oscillations
in the undersampled ``super-Nyquist'' regime of long-cadence (LC)
\emph{Kepler} data, i.e., \emph{above} the LC Nyquist frequency of
$\simeq 283\,\rm \mu Hz$. We conclude that the LC data may be utilized
for asteroseismic studies of targets whose dominant oscillation
frequencies lie as high as $\simeq 500\,\rm \mu Hz$. The frequency
detection threshold for the shorter-duration K2 science campaigns is
lower. We estimate that the robust threshold will lie somewhere
between $\simeq 400$ and $450\,\rm \mu Hz$.

These stars would usually be studied with \emph{Kepler} short-cadence
(SC) data. The oscillations are then oversampled, since the associated
Nyquist frequency of $\simeq 8496\,\rm \mu Hz$ lies well above the
frequencies of the detectable oscillations. However, the number of SC
available target slots is very limited: indeed, there are around only
50 targets of interest in the \emph{Kepler} archive that have detected
oscillations in SC data and $\nu_{\rm max}$ between the LC $\nu_{\rm
  Nyq}$ and $\simeq 500\,\rm \mu Hz$. In most of these cases there is
only 1\,month of data from the initial asteroseismic survey
phase. This precludes very detailed studies of these targets, which
demands high-frequency-resolution datasets for analyses of individual
modes (including rotational frequency splittings).

The exploitation of the archival \emph{Kepler} and new K2 data for LC
super-Nyquist studies offers the prospect of increasing significantly
the number of evolved subgiant and low-luminosity red-giant targets
available for asteroseismic study. By making use of the revised
catalogue of \emph{Kepler} star properties in Huber et al. (2014), we
estimate that up to 400 targets may be available for study in the
\emph{Kepler} archive. Moreover, many of these targets will have data
spanning the entire nominal Mission, giving the frequency resolution
needed for the abovementioned analyses of individual modes.

The potential number of targets amenable to study with K2 data could
run into the thousands, since new target lists will be generated for
each campaign.

\subsection*{ACKNOWLEDGEMENTS}

Funding for this Discovery mission is provided by NASA's Science
Mission Directorate. The authors wish to thank the entire
\emph{Kepler} team, without whom these results would not be possible.
W.J.C., Y.E., G.R.D., T.L.C., R.H. and A.M acknowledge the support of
the UK Science and Technology Facilities Council (STFC).  Funding for
the Stellar Astrophysics Centre is provided by The Danish National
Research Foundation (Grant agreement no.: DNRF106). The research
leading to these results has received funding from the European
Community's Seventh Framework Programme ([FP7/2007-2013]) under grant
agreement no. 312844 (SPACEINN).

\end{document}